\documentclass{llncs}
\usepackage{amssymb}
\usepackage{color}

\begin{document}

\title{Scope Logic with Local Reasoning and Pre/Post-State Properties
\thanks{This paper is supported}}
\author{ZHAO Jianhua, LI Xuandong}

\institute{State Key Laboratory of Novel Software Technology\\
    Dept. of Computer Sci. and Tech. Nanjing University\\
    Nanjing, Jiangsu, P.R.China 210093\\
    zhaojh@nju.edu.cn}
\maketitle

\newcommand{\seml}{[\![}
\newcommand{\semr}{]\!]}
\newcommand{\outlying}{\texttt{Outlying}}
\begin{abstract}
This paper presents an extension to Hoare logic for pointer program verification.
Logic formulas with user-defined recursive functions are used to specify properties on the program states before/after
program executions.

Three basic functions are introduced to represents  memory access, record-field access and
array-element access. Some axioms are introduced to specify these basic functions in our logic.

The concept Memory Scope Function (MSF) is introduced in our logic. Given a recursive function $f$,
the MSF of $f$ computes the set of memory units accessed during the evaluation of $f$. A set of rules
are given to derive the definition of this MSF syntactically from the definition of $f$.
As MSFs are also recursive functions, they also have their MSFs. An axiom is given to specify that
an MSF contains its MSF. Based on this axiom, local reasoning is supported with predicate variables.

Pre-state terms are used to specify the relations between pre-states and post-states. People can
use pre-state terms in post-conditions to represents the values on the pre-state.

The axiom of assignment statements in Hoare's logic is modified to deal with pointers. The basic idea
is that during the program execution, a recursive function is evaluated
to the same value as long as no memory unit in its memory scope is modified. Another proof rule is
added for memory allocation statements.

We use a simple example to show that our logic can deal with pointer programs in this paper. In the appendix,
the Shorre-Waite algorithm is proved using our logic. We also use the selection-sort program to show that
our logic can be used to prove program with indirectly-specified components.
\end{abstract}

\section{Introduction}
Hoare's  logic\cite{HOARE-1} can not deal with pointer programs because of pointer alias, i.e. many pointers may refer to the same location.
A few extensions to Hoare logic have been made to deal with pointers or shared mutable data structures \cite{RODNEY}\cite{STEPHEN}\cite{JOSEPH}.
Among them, separation logic \cite{SEPLOG} is famous.
That logic uses a memory model which consists of two parts: the stack and the heap. Pointers can only refer to data objects in the heap.
The Hoare logic is extended with a set of proof rules for heap lookup, heap mutation and program variable assignment.
Separation logic extends the predicate calculus with the separation-conjunction operator (*), which can separate the
heap into different disjoint parts.
To specify the separation conjunction and related function symbols($\mapsto$ and $\rightharpoondown$), totally 14 axioms
and proof rules are introduced. These axioms and rules are not strong enough to prove most programs. So five special classes of formulas are
defined based on the semantic of separation-conjunctions, $\mapsto$ and $\rightharpoondown$. 5 axioms and several semantic
theorems are introduced to improve Separation Logic. Yet another two new symbols ($\stackrel{[e]}{\hookrightarrow}$, $ \bigodot$) and  11 axioms are introduced to deal with composite types (record-types and array-types). Supplemental proof rules for assignment, allocation/deallocation are also provided. Another disadvantage of Separation Logic, people still have to deal with low-level address manipulations, which has been generally avoided in high-level programming languages. A few small programs have been used to demonstrate the potential of local reasoning for
scalability~\cite{SEPEXAMPLE}. In that paper, the Shorre-Waite algorithm was proved using Separation Logic, on the assumption of many complicated
logic formulas.

This paper presents an extension to Hoare logic for verification of pointer programs.
In order to deal with high-level program types abstractly and directly, we introduce three functions: $\ast$ (memory unit access),
$\&\rightarrow$ (record-field access), and $\&[\,]$ (array-element access). A set of axioms are used to model and specify the memory
layout/access in pointer programs.

The value stored in the memory unit referred by an address $x$ is $\ast x$. In this logic, the memory units for program variables and
heap objects are treated uniformly. A program variable $v$ corresponds to a constant memory address $\&v$. The value stored in $v$ is
expressed as $\ast\&v$.  As a consequence, program variables are no longer treated as logic variables in specifications.

Recursive functions are used in program specifications. For a recursive function defined using $\ast$, its value is relevant to a set
of memory units. We show that this memory unit set can be expressed by another recursive function, called memory-scope function (MSF). The definition
of the memory-scope function of $f$ can be constructed syntactically from the definition of $f$.

The concept pre-state terms are introduced in this paper. They can help people specify the relations between pre-/post-states.

This paper is organized as follows. We first introuce the syntax of programs and specifications in Section~\ref{SEC-PROG-SYNTAX}.
In Section~\ref{SEC-MEMSCOP-PRESTATE}, we introduce the concepts of memory scopes and pre-state terms.  A proof rule is introduced
to specify how definitions of MSFs are constructed. Two axioms are introduced to specify the properties about memory scopes.
To model memory access and layout in pointer programs, three functions and a set of axioms are introduced in Section~\ref{SEC-MEM-ACC-LAYOUT}.
The axioms and proof rules for program statements are given in Section~\ref{SEC-STATEMENTS}. Some other proof rules are given
in Section~\ref{SEC-OTHER-RULES}. Section~\ref{SEC-CONCISE-FORM} describes how to write verifications in 2-dimension form. The verification of the running example using our logic
is presented in Section~\ref{SEC-PROOF}. Section~\ref{SEC-CONCLUSION} concludes this paper.

Appendix~\ref{SEC-SHORRE-WAITE} presents the verification of an implementation of the Schorre-Waite algorithm. Appendix~\ref{SEC-SEL-SORT} presents an verification of a program template, i.e.
a program with indirectly-specified components.

\section{Syntax of programs and specifications}\label{SEC-PROG-SYNTAX}
In this section, we give a brief description of the small program language used in this paper.
\subsection{The syntax of program expressions}
Besides the basic arithmetic and logical operators, three operators are introduced: $\ast, \&[\,], \&\rightarrow n$. The operator $\ast$ is used to access the content stored in a memory unit; $\&[\,]$ is used to derive the address of an element of an array; $\&\,\rightarrow n$ is used to derive the address of a field of a record.

Formally, The syntax of program expressions is as follow.
\begin{enumerate}
\item For a program variable $v$, $\&v$ is an expression.
\item A constant (e.g. 1,2, \textbf{true}, \textbf{false}, \textbf{nil}) is an expression.
\item Let $e_0,e_1,e_2$ be expressions.
\begin{enumerate}
\item $e_0?e_1:e_2$ is also an expression;
\item $e_1+e_2$, $e_1-e_2$, $e_1*e_2$, $e_1/e_2$, $\textbf{not }e_1$, $e_1\textbf{ and }e_2$, $e_1\textbf{ or }e_2$ are expressions;
\item $\ast e_1$, $\&e_1[e_2]$, $\&e_1\rightarrow n$ are expressions.
\end{enumerate}
\end{enumerate}
If we compare this small language with the programming language C, the operators $\&\!\rightarrow n$ (or $\&[\,]$) can be viewed as a composition of $\&$ and $\rightarrow n$ (or $[\,]$ respectively). The semantic of $e_0?e_1:e_2$ is same as the conditional expression in C.

\subsection{The type system of the program language}
The small program language used in this paper is strong-typed.
In this paper, it is supposed that all programs under verification have passed static type check.
Each expression $e$ has a static type $t$, which means that at the runtime, either $e$ denotes a value of
type $t$ or $e$ is undefined.

The types used in programs include
the basic types: $\textbf{int}$ (for integers) and $\textbf{bool}$ (for boolean values); \emph{pointer-types} $\textbf{P}(t)$,
 \emph{array-types} $\textbf{ARR}(t, c)$, and \emph{record-types} $\textbf{REC}((n_1,t_1)\times \dots\times (n_k, t_k))$, where
$t,t_1,\dots, t_k$ are types, $n_1,n_2,\dots,n_k$ are $k$ different names, $c$ is a positive integer constant.
We use $\textbf{Ptr}$ as the super type of all pointer-types. People can also define user-types using the form $name:=type$.

If we view the new operators $\&\rightarrow$ and $\&[]$ as compositions of the C operator $\&$, $\rightarrow$ and $[\,]$, the type rules and well-formed conditions of program expressions are similar to that of the programming language C. However, it is required that
$\ast$ can only be applied to values of the set $\textbf{P}(\textbf{int}), \textbf{P}(\textbf{bool})$ and $\textbf{P}(\textbf{P}(t))$ for some $t$.

\subsubsection{Abbreviations}
Using the operators $\ast, \&[\,]$ and $\&\!\rightarrow n$, we can fulfill any operation on arrays and records. However,
an expression using $\&\!\rightarrow n$ and $\&[\,]$ directly may be difficult to write/read. For conciseness, we can use the following abbreviations in the programs such that we can write C-like expressions. (A new operator $(\&.)$ is introduced here.)
\begin{itemize}
\item $\ast (\&v)$, $\ast (\&e\rightarrow n)$, $\ast (\& e. n)$,  and  $\ast(\& e_1[e_2])$ can be abbreviated as $v$, $e\rightarrow n$, $e.n$,  and $e_1[e_2]$ respectively.
\item $\&(\&v)\!\rightarrow n$, $\&(\&e_1\rightarrow n_1)\!\rightarrow n_2$, $\&(\&e_1.n_1)\!\rightarrow n_2$ and $\&(\&e_1[e_2])\!\rightarrow n$ can be abbreviated as $\&v.n$, $\&(e_1\rightarrow n_1).n$, $\&(e_1.n_1).n_2$ and $\&e_1[e_2].n$ respectively.

\item $\&(\&v[e_2])[e_3]$, $\&(\&e_1\rightarrow n)[e_3]$, $\&(\&e_1.n)[e_3]$, and  $\& (\&e_1[e_2])[e_3]$ can be abbreviated as $\&(v[e_2])[e_3]$, $\&(e_1\rightarrow n)[e_3]$, $\&(e_1.n)[e_3]$ and $\&(e_1[e_2])[e_3]$ respectively.
\end{itemize}
We can write concise expressions using these abbreviation rules.

\begin{example}
Let $a$, $i$ and $j$ be three program variables. The type of $i$ and $j$ is $\textbf{int}$; The type of $a$ is $\textbf{ARR}(\textbf{ARR}(\textbf{REC}((f1,\textbf{int})\times(f2,\textbf{bool})), 100),100)$.
We can write $\ast(\&(\&(\&(\&a)[\ast(\&i)])[\ast(\&j)])\rightarrow f1)$ to access the field $f1$ of the element in the $i$th row and the $j$th column.
Using the abbreviation rules above, we can abbreviate this expression as
$a[i][j].f1$.
\end{example}

\subsection{The syntax of program statements}
The syntax of program statements is as follows.
$$\begin{array}{rcl}
st & ::=\ \ &\texttt{skip}\ \ |\ \ \ast e_1:=e_2\ \ |\ \ \ast e:=\texttt{alloc}(t)\ \ \\
   &        &|\ \ st;\ st\ \ |\ \ \textbf{if}\ (e) \ st\ \textbf{else}\ st\ \ |\ \ \textbf{while}\ (e)\ st\\
\end{array}$$
The type-rules and well-conditions of the statements are as follow.
\begin{itemize}
\item For the assignment $\ast e_1:=e_2$, $\ast e_1$ and $e_2$ must have same static type. Further more, their type must \textbf{integer}, \textbf{boolean}, or a pointer type.
\item For the memory allocation statement $\ast e:=\texttt{alloc}(t)$, the static type of $\ast e$ must be $\textbf{P}(t)$.
\item For the \textbf{while}-statement and \textbf{if}-statement, the static type of $e$ must be $\textbf{boolean}$.
\end{itemize}
The operational semantic of these statements is similar to that of the programming language C.

\begin{example}
The program depicted in Figure~\ref{EXAMPLE} is a running example
used in this paper. The type of $\textsf{k}$ and
$\textsf{d}$ is \textbf{int}. The type of $\textsf{root}$ and $\textsf{p}$ is
\textbf{P}($\texttt{Node}$), where $\texttt{Node}=\textbf{REC}((l, \textbf{P}(\texttt{Node}))\times(r,\textbf{P}(\texttt{Node}))\times (K,\textbf{int})\times(D,\textbf{int}))$. This program
first searches a binary search tree for a node of which the field $K$ equals $\textsf{k}$. Then it sets the
filed $D$ of this node to $\textsf{d}$.
\end{example}

\begin{figure}
\begin{center}
\parbox{300pt}{
\begin{tabbing}
\textsf{pt}:=\textsf{root};\\
\textbf{while} \= ($\textsf{pt}\!\rightarrow\!\!K \neq \textsf{k}$)\\
\{\\
\>\textbf{if} ($\textsf{k} < \textsf{pt}\rightarrow K$ ) $\textsf{pt} := \textsf{pt}\rightarrow l$ \textbf{else} $\textsf{pt} := \textsf{pt}\rightarrow r$;\\
 \}\\
$\textsf{pt}\rightarrow D := \textsf{d}$;
\end{tabbing}}
\end{center}
\caption{The program used as a running example}\label{EXAMPLE}
\end{figure}

\subsection{Syntax of specifications}\label{SEC-SYN-SPEC}
In our logic, a specification is a Hoare-Triple of the form
$$p\ \{\ c\ \}\ q$$
where  $c$ is a code fragment, $p$ and $q$ are logic formulas with recursive functions. The formulas $p$ and $q$ are respectively called  the pre-condition and post-condition of this specification.
As we will use recursively-defined partial functions in program specifications, $p$ and $q$ should be formulas of some three-value logics that can deal with undefinedness. In this paper, we use the Logic for Partial Functions (LPF) \cite{LPF} as the basic logic, i.e. $p$ and $q$ are LPF formula. However, other three-value logics can also be used.

There also some other things should be noticed here.
\begin{itemize}
\item Program variables and logic variables are different things in our logic. A program variable represents a memory block storing values. The values stored in a program variable may changed by code. A logic variable is a place-holder for a value. A logic variable appeared in pre-condition and post-condition represents the same value.
\item In our logic, the function $\ast$ is used to represents the program state. So $\ast$ has different interpretations in $p$ and $q$. Consequently, the recursive functions in the specification defined based on $\ast$ also have different interpretations accordingly.
\end{itemize}
 Formulas appeared in specifications are also strong-typed. Each logic variable has a static type, though we may omit it. The static types of terms in such formulas can be automatically decided.

\begin{example}
Let
 $\textsf{isHBST}$, $\textsf{Dom}$, $\textsf{Map}$ and $\textsf{FldD}$ be the recursive functions defined in Figure~\ref{DATA-STRUCTURE-INTERPRETATION-FUNCTIONS};
 $Prog$ be the program depicted in Figure~\ref{EXAMPLE}.
The specification
\begin{equation}\begin{array}{l}
\texttt{Outlying}(\rho,\{\&\textsf{p}\}\cup
\textsf{FldD}(\textsf{root})\})\land
\textsf{isHBST}(\textsf{root})\land \textsf{k} \in
\textsf{Dom}(\textsf{root})\\
\mbox{}\ \ \ \ \ \ \ \ \ \  \{\ \ Prog\ \ \}\\
\rho\land \textsf{isHBST}(\textsf{root})\land
\textsf{Map}(\textsf{root})=\overleftarrow{\textsf{Map}(\textsf{root})}\dag\{\textsf{k}\mapsto
\textsf{d}\}
\end{array}\label{WHOLE-SPECIFICATION}\end{equation}
says that if the program starts it execution when $\textsf{root}$ points to a binary search and $\textsf{k}$ equals to the field $K$ of some node in this tree, $\textsf{root}$ still points to a binary search tree after the execution. The finite map represented by this tree
is the same as the original map except that it maps $\textsf{k}$ to $\textsf{d}$.

$\overleftarrow{\textsf{Map}(\textsf{root})}$ in the post-condition is a pre-state term. It means the map represented by the BST on the pre-state. The definition and proof rules of pre-state terms will be presented later.

$\texttt{Outlying}(\rho,\{\&\textsf{p}\}\cup
\textsf{FldD}(\textsf{root})\})$ is an abbreviation for $\rho\land (\mathfrak{M}(\rho)\cap (\{\&\textsf{p}\}\cup
\textsf{FldD}(\textsf{root}))\}=\emptyset)$. $\mathfrak{M}(\rho)$ denotes the memory scope of the formula $\rho$.
This specification shows that the program modifies only the memory units in $\{\&\textsf{p}\}\cup
\textsf{FldD}(\textsf{root})$. If $\rho$ holds on the pre-state, and is irrelevant to the memory units in this set, $\rho$ still holds on the post-state.  The concept of memory scopes will be explained in the next section.

\end{example}

\begin{figure}
\begin{center}
\begin{tabbing}
$\textsf{NodeSet}(x:\textbf{P}(\texttt{Node})):\textbf{SetOf}(\textbf{Ptr})$\\
\ \ \ \ \ \ \ \ \=$\triangleq$ $(x=\textbf{nil}) ? $\=\ $\emptyset : (\{x\}\cup \textsf{NodeSet}(x\rightarrow l)\cup \textsf{NodeSet}(x\rightarrow r))$\\
\\
$\textsf{Map}(x:\textbf{P}(\texttt{Node})): \textbf{Map}(\textbf{int}, \textbf{int})$\\
        \>$\triangleq(x=\textbf{nil}) ? \emptyset :\{x\rightarrow K\mapsto x\rightarrow D\}\dag \textsf{Map} (x\rightarrow l)\dag \textsf{Map}(x\rightarrow r)$\\
\\
$\textsf{MapP}(x:\textbf{P}(\texttt{Node}),y:\textbf{P}(\texttt{Node})):\textbf{Map}(\textbf{int}, \textbf{int})$\\
        \>$\triangleq(x=\textbf{nil}) ? \emptyset : \textsf{MapP}(x\rightarrow l) \dag \textsf{MapP} (x\rightarrow r)\dag$\\
           \>\ \ \ \ \ \ \ $((x = y)? \emptyset : \{x\rightarrow K\mapsto x\rightarrow D\})$\\
\\
$\textsf{Dom}(x:\textbf{P}(\texttt{Node})):\textbf{SetOf}(\textbf{integer})$\\
        \>$\triangleq (x=\textbf{nil}) ? \emptyset : (\{x \rightarrow K\} \cup \textsf{Dom}(x\rightarrow l)\cup \textsf{Dom}(x\rightarrow r))$\\
\\
$\textsf{isHBST}(x:\textbf{P}(\texttt{Node})): \textbf{boolean}$\\
        \>$\triangleq (x=\textbf{nil}) ? \textbf{true} : \texttt{InHeap}(x)\land \textsf{isHBST}(x\rightarrow l)\land \textsf{isHBST}(x\rightarrow r)\land$\\
         \>         \> $(\textsf{Dom}(x\rightarrow l)=\emptyset?\texttt{true}: \texttt{MAX}(\textsf{Dom}(x\rightarrow l))<x\rightarrow K) \land $\\
         \>         \>$(\textsf{Dom}(x\rightarrow r)=\emptyset?\texttt{true}:x\rightarrow K < \texttt{MIN}(\textsf{Dom}(x\rightarrow r)))$\\
\\
$\textsf{FldD}(x:\textbf{P}(\textsf{Node})):\textbf{SetOf}(\textbf{Ptr})$\\
       \>$\triangleq (x=\textbf{nil})? \emptyset : (\{\& x\rightarrow
D\}\cup \textsf{FldD}(x\rightarrow l)\cup \textsf{FldD}(x\rightarrow
r))$
\end{tabbing}
\end{center}
\caption{The definitions of a set of recursive functions}\label{DATA-STRUCTURE-INTERPRETATION-FUNCTIONS}
\end{figure}

\section{Memory scopes and pre-state terms}\label{SEC-MEMSCOP-PRESTATE}

\subsection{Terms denoting pre-state values}
In many cases, people are interested in the relationship between the values before/after program executions.
Merely specifying the property of the program state after program execution is not sufficient for this purpose.
In our logic, we use \emph{pre-state terms} to denoting values on pre-states in post-conditions.
\begin{definition}
\emph{\textbf{Pre-state terms}} Let $e$ be a term containing no pre-state sub-term, $\overleftarrow e$ is a pre-state term.
\end{definition}
Pre-state terms can appear in both pre-conditions and post-conditions. A pre-state term $\overleftarrow e$ denoting the value of $e$ interpreted based on the pre-state before execution. In pre-conditions, $\overleftarrow e$ and $e$ is interchangeable.
We have the following two proof rules.
$$\frac{p[e/x]\ \{s\}\ q}{p[\overleftarrow{e}/x]\ \{s\}\ q}\ \ \ \ (\textrm{PRE-1})\ \ \ \ \ \ \ \ \ \ \ \ \ \ \ \ \frac{p[\overleftarrow{e}/x]\ \{s\}\ q}{p[e/x]\ \{s\}\ q}\ \ \ \ (\textrm{PRE-2})$$
In the rest of this paper, we will inter-change $e$ and $\overleftarrow e$ in pre-conditions without mentions these proof rules.
For convenience, for a formula $Q$ (or a term $e$) containing pre-state terms, we use $\overline{Q}$ (or $\overline e$ respectively) to denote the formula (or term) derived by replacing all pre-state terms in $Q$ (or $e$) with their original forms. For example, $\overline{\overleftarrow{\ast \textsf{pt}}=\textsf{t}}$ is $\ast \textsf{pt}=\textsf{t}$.

\subsection{Memory scopes}
\subsubsection{Memory scope form of terms}
An execution of a piece of code may change the values stored in some memory units.
The function symbol $\ast$ has different interpretations in the pre-condition and post-condition of a specification.
So does the functions defined based on $\ast$. As a consequence, a term may denote different values
in the pre-condition and the post-condition.
However, the value of a term $e$ relies only on a finite set of memory units. This set can also be
expressed as a term, called the memory scope form of $e$, denoted as $\mathfrak{M}(e)$.
Given a term $e$,  $\mathfrak{M}(e)$ is defined as follow.
\begin{itemize}
\item For any pre-state term $\overleftarrow e$, $\mathfrak{M}(\overleftarrow e)$ is $\emptyset$.
\item If $e$ is a logical variable or a constant, $\mathfrak{M}(e)$ is $\emptyset$.
\item If $e$ is of the form $f(e_1,\dots,e_n)$, $\mathfrak{M}(e)$ is $\mathfrak{M}(e_1)\cup\dots\cup \mathfrak{M}(e_n)\cup \mathfrak{M}(f)(e_1,\dots,e_n)$, where $\mathfrak{M}(f)$ is the memory scope function of $f$.
\item If $e$ is of the form $e_0 ? e_1: e_2$,  $\mathfrak{M}(e)$ is $\mathfrak{M}(e_0)\cup(e_0? \mathfrak{M}(e_1) :\mathfrak{M}(e_2))$.
\end{itemize}
Though $\overleftarrow e$ is evaluated on the pre-state, its value is not modified by the execution. So, $\mathfrak{M}(\overleftarrow e)$ is $\emptyset$.

\subsubsection{Memory scope functions}
Intuitively speaking, $\mathfrak{M}(f)(x_1,\dots,e_x)$ computes the set of memory units accessed during the evaluation of $f(x_1,\dots,x_n)$.
The argument types of $\mathfrak{M}(f)$ is the same as those of $f$. The result type of $\mathfrak{M}(f)$ is $\textbf{SetOf}(\textbf{Ptr})$.
Given a function $f$, the definition of $\mathfrak{M}(f)$ can be derived as follow.
\begin{itemize}
\item $\mathfrak{M}(f)(x_1,\dots,x_n)\triangleq \emptyset$ if $f$ is a function symbol associated with basic types or abstract types (for example, $+,-,\times, /, >, <, \in, \subseteq \dots$), or $\&\rightarrow n$, $\&[\,]$, $\&v$ for some program variable, $\textbf{nil}$.
\item $\mathfrak{M}(\ast)(x) \triangleq \{x\}$. That is, $\ast x$ access the memory unit referred by $\{x\}$.
\item If $f$ is (recursively) defined as $f(x_1,\dots, x_n)\triangleq e$, we have the definition $\mathfrak{M}(f)(x_1,\dots,x_n)\triangleq\mathfrak{M}(e)$.
\end{itemize}
In LPF, a function definition is also a formula. We have the following proof rule.
$$\frac{\ \ \ \ \ \ f(x_1, \dots, x_n)\triangleq e}{\ \ \ \ \ \ \ \mathfrak{M}(f)(x_1, \dots, x_n)\triangleq\mathfrak{M}(e)}\ \ \ \ \ \ \ \ \ \ \textrm{(SCOPE-FUNC)}$$

\begin{example}
Let $e$ be $a[i][j].f1$ (i.e. $\ast(\&(\&(\&(\&a)[\ast(\&i)])[\ast(\&j)])\rightarrow f1)$), $\mathfrak{M}(e)$ is $\{\&i, \&j, \&(a[i][j].f1)\}$.

The memory scope form of $(x=\textbf{nil}) ? \emptyset : (\{x\}\cup \textsf{NodeSet}(x\rightarrow l)\cup \textsf{NodeSet}(x\rightarrow r))$ is equivalent to $(x=\textbf{nil}) ? \emptyset : \{\&x\rightarrow l\}\cup \mathfrak{M}(\textsf{NodeSet})(x\rightarrow l)\cup \{\&x\rightarrow r\}\cup\mathfrak{M}(\textsf{NodeSet})(x\rightarrow r)$. According to the definition of  $\textsf{NodeSet}$ in Figure~\ref{DATA-STRUCTURE-INTERPRETATION-FUNCTIONS},
$$\begin{array}{l}
\mathfrak{M}(\textsf{NodeSet})(x)\triangleq (x=\textbf{nil}) ? \emptyset : \{\&x\rightarrow l, \&x\rightarrow r\}\cup\\
\mbox{}\ \ \ \ \ \ \ \ \ \ \ \ \mathfrak{M}(\textsf{NodeSet})(x\rightarrow l)\cup \mathfrak{M}(\textsf{NodeSet})(x\rightarrow r)
\end{array}$$
The definition of $\mathfrak{M}(\textsf{NodeSet})$ is equivalent to that of $\textsf{MSF}_{lr}$ depticted in Figure~\ref{SCOPE-FUNCTIONS}
The following table lists the memory scope functions for the functions defined in Figure~\ref{DATA-STRUCTURE-INTERPRETATION-FUNCTIONS} and Figure~\ref{SCOPE-FUNCTIONS}.
\begin{center}
\begin{tabular}{|c|c|}
\hline
Recursive functions  &  Memory Scope Functions\\
\hline
\textsf{NodeSet}, $\textsf{MSF}_{lr}$, $\textsf{MSF}_{lrk}$, $\textsf{MSF}_{lrkd}$, $\textsf{FldD}$     &       $\textsf{MSF}_{lr}$\\
\hline
\textsf{Dom}, \textsf{isHBST}  &   $\textsf{MSF}_{lrk}$\\
\hline
\textsf{Map}             &   $\textsf{MSF}_{lrkd}$\\
\hline
\textsf{MapP}            &   $\textsf{MPP}_m$\\
\hline
\end{tabular}
\end{center}
\hfill$\Box$
\end{example}

\begin{figure}
\begin{tabbing}
$\textsf{MSF}_{lr}(x:\textbf{P}(\textsf{Node})): \textbf{SetOf}(\textbf{Ptr})$\\
\ \ \ \ \ \ \ \ \=$\triangleq$ $(x=\textbf{nil}) ? $\=\ $\emptyset : (\{\&x\rightarrow l, \&x\rightarrow r\}\cup \textsf{MSF}_{lr}(x\rightarrow l)\cup \textsf{MSF}_{lr}(x\rightarrow r))$\\
\\
$\textsf{MSF}_{lrk}(x:\textbf{P}(\textsf{Node})): \textbf{SetOf}(\textbf{Ptr})$\\
        \>$\triangleq (x=\textbf{nil}) ? \emptyset : (\{\&x \rightarrow K, \&x\rightarrow l, \&x\rightarrow r\} \cup \textsf{MSF}_{lrk}(x\rightarrow l)\cup\textsf{MSF}_{lrk}(x\rightarrow r))$\\
\\
$\textsf{MSF}_{lrkd}(x:\textbf{P}(\textsf{Node})): \textbf{SetOf}(\textbf{Ptr})$\\
        \>$\triangleq(x=\textbf{nil}) ? \emptyset :\{\&x\rightarrow K, \&x\rightarrow D, \&x\rightarrow l, \&x\rightarrow r\}\cup \textsf{MSF}_{lrkd}(x\rightarrow l)\cup \textsf{MSF}_{lrkd}(x\rightarrow r)$\\
\\

$\textsf{MPP}_m(x:\textbf{P}(\textsf{Node}), y:\textbf{P}(\textsf{Node}) ): \textbf{SetOf}(\textbf{Ptr})$\\
        \>$\triangleq(x=\textbf{nil}) ? \emptyset :\{\&x\rightarrow l, \&x\rightarrow r\}\cup \textsf{MPP}_m(x\rightarrow l) \cup \textsf{MPP}_m (x\rightarrow r)\cup$\\
           \>\ \ \ \ \ \ \ $((x = y)? \emptyset : \{\&x\rightarrow K, \&x\rightarrow D\})$\\
\end{tabbing}
\caption{The scope functions}\label{SCOPE-FUNCTIONS}

\end{figure}

\subsubsection{Memory Scope for Formulae}
Roughly speaking, the memory scope term of a formula is the union of the memory scopes of the terms of this formula.
To formally define the memory scope of a formula $p$, it is required that each quantified variable in $p$ has a unique
name, and is different from the free variables in $p$. Please be noticed that for an arbitrary formula $p$, we can easily get an equivalent formula satisfying this condition by variable renaming. The memory scope term of a formula $p$ is defined as follow.
\begin{enumerate}
\item If $p$ is a boolean-typed term $e$, $\mathfrak{M}(p)$ is $\mathfrak{M}(e)$.
\item If $p$ is $e_1=e_2$ or $e_1==e_2$, $\mathfrak{M}(p)$ is $\mathfrak{M}(e_1)\cup \mathfrak{M}(e_2)$. The symbol $==$ is the strong equivalence in LPF. It means that either both side of $==$ are undefined, or they equal to each other.
\item If $p$ is $p_1\land p_2$, $\mathfrak{M}(p)$ is $\mathfrak{M}(p_1)\cup \mathfrak{M}(p_2)$.
\item If $p$ is $\neg p'$, $\mathfrak{M}(p)$ is $\mathfrak{M}(p')$
\item If $p$ is $\forall x\cdot p'$, $\mathfrak{M}(p)$ is $\mathfrak{M}(p')$.
\end{enumerate}

\subsubsection{An axiom about memory scopes}
If all the functions in a term $e$ are either basic functions or recursively defined functions,  we can find that the evaluation of $\mathfrak{M}(e)$ tracks the evaluation process of $e$, and records all the memory units accessed during this process.
All the memory units accessed during the evaluation of $\mathfrak{M}(e)$ are also accessed during the evaluation of $e$. We have the following two axioms about memory scope functions.

$$\mathfrak{M}(p)\!\downarrow\ \Rightarrow \mathfrak{M}(\mathfrak{M}(p))\subseteq \mathfrak{M}(p)\ \ \ \ \ \ \ \ \ \ \ \ (\textrm{SCOPE-1})$$

$$\mathfrak{M}(e)\!\downarrow\ \Rightarrow \mathfrak{M}(\mathfrak{M}(e))\subseteq \mathfrak{M}(e)\ \ \ \ \ \ \ \ \ \ \ \ (\textrm{SCOPE-2})$$
Here, $\downarrow$ is an operator in LPF. $\mathfrak{M}(e)\downarrow$ and $\mathfrak{M}(p)\downarrow$ respectively mean that $\mathfrak{M}(e)$ and $\mathfrak{M}(p)$ are denoting.

\section{Supporting local reasoning}
In our logic, local reasoning is supported using predicate variables.
Usually, a specification is of the form
$$\rho\land (\mathfrak{M}(\rho)\cap e =\emptyset )\land pre\ \ \{\ s\ \}\ \ \rho\land post$$
We can substitute $\rho$ with other formulas to get new assertions about $s$.
The static type of the term $e$ is $\textbf{SetOf}(\textbf{Ptr})$. It is an over-approximation of the set of memory units modified by the code $s$. The premise $(\mathfrak{M}(\rho)\cap e =\emptyset)$ means that $\rho$ is irrelevant to the values stored in the memory units in $e$. Thus, $\rho$ still holds when $s$ halts normally.

For conciseness, in the rest of this paper, we use $\outlying(\rho,e)$ as an abbreviation for $\rho\land (\mathfrak{M}(\rho)\cap e =\emptyset)$. The assertion above can be written as
$$\outlying(\rho, e)\land pre\ \ \{\ s\ \}\ \ \rho\land post$$

\begin{example}
Considering the program depicted in Figure~\ref{EXAMPLE}. The program
changes only the memory unit $\&\textsf{p}$ and the field $D$ of some
node of the tree. The specification~\ref{WHOLE-SPECIFICATION} shows that for any predicate $\rho$ which is irrelevant to
the memory unit $\&\textsf{p}$ and
the field $D$ of any nodes in the tree, if $\rho$ holds on the pre-state, it still holds on the post-state.
\end{example}

\begin{theorem}\label{THEOREM-EXPANDING}
Let $\rho$ be an arbitrary predicate symbol. Let $r$ be a predicate. As mentioned previously, $\overline{r}$ is derived by replacing each pre-state term $\overleftarrow{e}$ with $e$.
$$\frac{ \mathtt{Outlying}(\rho,e)\land p\ \{\ s\ \}\ \rho\land q\ \ \ \ \ \ p \land\overline{r}\Rightarrow (\overline{\mathfrak{M}(r)}\cap e=\emptyset) }
{p \land \overline{r}\ \ \{\ s\ \}\ \ q\land r}$$
\end{theorem}
\begin{proof}
Substitute $\rho$ with $r$ and expand $\outlying$ in the first premise, we have
$$r\land (\mathfrak{M}(r)\cap e=\emptyset)\land p\ \{\ s\ \}\ r\land q$$
Apply the proof rule \textrm{PRE-2}, we have
$$\overline{r}\land (\overline{\mathfrak{M}(r)}\cap e=\emptyset)\land p\ \{\ s\ \}\ r\land q$$
From the second premise and the consequence rule (presented later), we have
$$\overline{r}\land p\ \{\ s\ \}\ r\land q$$
\hfill$\square$
\end{proof}
Theorem~\ref{THEOREM-EXPANDING} will be used frequently to derive global assertions based on the local assertion of a statement. We call an application of
this theorem as an \emph{expanding}.


\section{The axioms about memory access and layout}\label{SEC-MEM-ACC-LAYOUT}
In this section, we present some axioms about memory access and memory layouts of composite types.
In the implementation of real program languages, memory layouts for composite types are compiler-dependent.
So computing the address of a component in a composite-typed data by address-offsetting is
not a good choice. In our logic, we use a set of axioms to specify such memory layouts abstractly.

\subsection{The auxiliary function \texttt{Block}}
We define an auxiliary function $\texttt{Block}:\textbf{Ptr}\rightarrow \textbf{setof}(\textbf{Ptr})$ to denote the set of memory units in a memory block. The definition of $\texttt{Block}$ is as follows.

\noindent
$\texttt{Block}(r)= \emptyset$ if $r=\textbf{nil}$. Otherwise $\texttt{Block}(r)=$
$$\left\{
\begin{array}{rcl}
\{r\}  &\ \   & \mbox{if $\ast r$ is of type $\textbf{int}$, $\textbf{bool}$ or $\textbf{Ptr}$}\\
\bigcup_{\mbox{\tiny $n$: field name of $t$}}\texttt{Block}(\&r\rightarrow n)&\ \
&\mbox{if $r:\textbf{P}(t)$ and $t$ is a record type}\\
\bigcup_{i=0}^{c-1} \texttt{Block}(\&r[i]) &\ \ &\mbox{if $r:\textbf{P}(\textbf{ARR}(t',c))$ for some $t'$}
\end{array}
\right. $$
Intuitively speaking, $\texttt{Block}(r)$ is the set of memory units in the memory block referred by $r$.

\subsection{The axioms}
There is no memory-deallocation statement in our small program language.
Each time a memory block is allocated, the memory units of pointer types in this block are initialized to $\textbf{nil}$. Thus, a non-nil pointer always refers to a valid memory unit or memory block. We have the following axiom.
$$x:\textbf{P}(t)\land x\neq \textbf{nil}\Rightarrow \ast x:t\ \ \ \ \ \ \ \ \ \ (\textrm{MEM-ACC})$$
According to the type rule of $\ast$, $t$ can only be $\textbf{int}$, $\textbf{bool}$, or pointer types. The predicate $e:t$ means that the term $e$ denotes a value of the type $t$.

Two different memory blocks are either disjoint, or one of them contains another one. We have the following axiom.
$$x\neq y\Rightarrow \begin{array}{c}\texttt{Block}(x)\cap
\texttt{Block}(y)=\emptyset\lor\\
\texttt{Block}(x)\subset \texttt{Block}(y) \lor
\texttt{Block}(y)\subset \texttt{Block}(x)\end{array}\ \ \ \ \ \ \ \ \ \ (\textrm{MEM-BLK})$$
Here, the static type of $x$ and $y$ must be pointer types.

A unique memory block is assigned to each declared program variable. Furthermore, such a memory block is not contained by any other
memory blocks. So we have the following axioms.
$$\begin{array}{lr}
\& v\neq \textbf{nil} &(\textrm{PVAR-1})\\
\&v_1\neq \&v_2 &(\textrm{PVAR-2})\\
\texttt{Block}(\&v)\not\subset \texttt{Block}(x)\ \ \ \ \ \ \ \ \ \ \ \ &(\textrm{PVAR-3})\end{array}$$
Here $v$ is a program variable, $v_1$ and $v_2$ are two different program variables. The static type of $x$ is a pointer type.

A memory block for a record-typed data is allocated as a whole. That is, when a record-typed memory block is allocated,  the blocks for all its fields are allocated. We have the following axiom.
$$x\neq \textbf{nil}\Rightarrow \&x\rightarrow n\neq \textbf{nil}\ \ \ \ \ \ \ \ \ \ (\textrm{REC-1})$$
Here, the static type of $x$ is $\textbf{P}(\textbf{REC}(\dots\times(n,t)\times\dots))$

In a record-typed memory block, the memory blocks for different fields are disjoint with each other. If the static type of $x$ is
$\textbf{P}(\textbf{REC}(\dots\times(n_1,t_1)\times\dots\times(n_2,t_2)\times\dots))$, we have the following axiom.
$$x\neq \textbf{nil}\Rightarrow \texttt{Block}(\&x\rightarrow n_1)\cap \texttt{Block}(\&x\rightarrow
        n_2)=\emptyset\ \ \ \ \ \ \ \ \ \ (\textrm{REC-2})$$

Similarly, we have the following two axioms. Here, the static type of $x$ is $\textbf{P}(\textbf{ARR}(t,c))$. The static type of $y, y_1, y_2$ is \textbf{int}.
The axiom \textrm{ARR-1} says that an array-typed memory block is allocated as a whole, i.e. when an array-typed memory block is allocated, all of the memory blocks for its elements are allocated. The axiom \textrm{ARR-2} says that the memory blocks allocated for different elements are disjoint with each other.
$$(x\neq \textbf{nil})\land (0\le y< c) \Rightarrow \&x[y]\neq \textbf{nil}\ \ \ \ \ \ \ \ \ (\textrm{ARR-1})$$
$$\begin{array}{c}
(x\neq \textbf{nil})\land (0\le y_1 < c)\land (0\le y_2<c) \land (y_1\neq y_2) \Rightarrow\\
\mbox{}\ \ \ \ \ \ \ \ \ \ \ \ \ \ \ \ \ \ \ \ \ \texttt{Block}(\&x[y_1])\cap \texttt{Block}(\&x[y_2])=\emptyset
\end{array}\ \ \ \ \ \ \ \textrm{(ARR-2)}$$

\section{Axioms and proof rules of program statements}\label{SEC-STATEMENTS}
In this section, we present the axioms and proof rules to specify the effect of program statements.
There are three axioms for primitive statements and three proof rules for control flow statements.

\subsection{The axiom for \texttt{skip}}
The skip statement changes nothing. If a formula $q$ hold before the execution of $\texttt{skip}$, $q$ still holds after the execution.
This is specified by the following axiom.
$$q\ \{\ \texttt{skip}\ \}\ q\ \ \ \ \ \ \ \ \ \ (\textrm{SKIP-ST})$$

\subsection{The axiom for assignment}
For an assignment $\ast e_1 := e_2$, it is required that $e_1$ evaluates to a non-nil pointer, i.e. it refers to a memory unit,
and $e_2$ evaluates to a value on the pre-state. One the post-state, the memory unit stores the value of $e_2$ evaluated on the pre-state.
Furthermore, if $e_1$ is not in the memory scope of a formula $\rho$, $\rho$ still holds after the execution. This is specified by the following axiom.
$$\begin{array}{l}
\outlying(\rho, \{e_1\}) \land(e_1\neq\textbf{nil})\land(e_2\downarrow)\\
\mbox{}\ \ \ \ \ \{\ast e_1 := e_2\}\\
\rho\land (\ast \overleftarrow {e_1} =  \overleftarrow {e_2})\end{array}\ \ \ \ (\textrm{ASSIGN-ST})$$
Here, $\downarrow$ is an operator in LPF: $e_2\downarrow$ means that $e_2$ is evaluated to some value.

\begin{example}\label{ASSIGN-ASSERTIONS}
Considering the assignment $\textsf{pt} := \textsf{pt}\rightarrow l$ in the program depicted in Figure~\ref{EXAMPLE}.  As $
\&\textsf{pt}\neq \textbf{nil}$ and $(\textsf{pt}\neq \textbf{nil} )\Rightarrow (\&\textsf{pt}\rightarrow l\neq \textbf{nil}) \Rightarrow(\textsf{pt}\rightarrow l)\downarrow$, we can have the following assertion.
$$\outlying(\rho, \{\&\textsf{pt}\})\land(\textsf{pt}\neq\textbf{nil})\ \{\textsf{pt} := \textsf{pt}\rightarrow l\}\ \rho\land (\ast \overleftarrow {\& \textsf{pt}} =  \overleftarrow {\textsf{pt}\rightarrow l})$$

Because $\mathfrak{M}(\&\textsf{pt}) \cap\{ \&\textsf{pt}\} = \emptyset$, substitute $\rho$ with $\rho\land (\overleftarrow{\&\textsf{pt}} =\&\textsf{pt})$, and apply the consequence rule (presented later), we have
$$\begin{array}{l}\outlying(\rho, \{\&\textsf{pt}\})\land(\textsf{pt}\neq\textbf{nil})\ \{\textsf{pt} := \textsf{pt}\rightarrow l\}\ \rho\land (\textsf{pt} =  \overleftarrow {\textsf{pt}\rightarrow l})
\end{array}$$
\end{example}

\subsection{The axiom for memory allocation}

For an allocation statement $\ast e_1 := e_2$, it is required that $e_1$ evaluates to a non-nil pointer. After the execution,
the value stored in the memory unit referred by $\overleftarrow{e_1}$ refers to a memory block that is unreachable on the pre-state.
This allocation statement modifies only the memory unit referred by $e_1$ and the memory unit newly allocated. If a predicate $\rho$ holds on the pre-state and $e_1$ is not in the memory scope of $\rho$, $\rho$ still holds on the post-state. This is specified by the following axiom.
$$\begin{array}{l}
\outlying(\rho,\{e_1\}) \land (e_1\neq \textbf{nil}) \\
\mbox{}\ \ \ \ \ \ \ \ \ \ \{*e_1:=\textsf{alloc}(t)\}\\
\outlying(\rho, \texttt{Block}(\ast \overleftarrow{e_1}))\land(\ast\overleftarrow{e_1}\neq \textbf{nil})\land \texttt{Init}(\ast \overleftarrow{e_1})
\end{array}\ \ \ \ (\textrm{ALLOC-ST}) $$
The predicate $\texttt{Init}(x)$ is defined as $\forall y\in \texttt{Block}(x)\cdot y:\textbf{P}(\textbf{P}(t)) \Rightarrow \ast y=\textbf{nil}$. It means that all the pointers stored in the block referred by $x$ is initialized to \textbf{nil}.
%

\begin{example}
Considering the sequential statements $\textsf{t}:=\textsf{alloc}(\textsf{Node});\ \ \textsf{t}\rightarrow k := \textsf{k};\ \ \textsf{t}\rightarrow d:=\textsf{d}$.
From the axiom \textrm{ALLOC-ST}, we have
$$\begin{array}{l}
\outlying(\rho,\{\&\textsf{t}\})\\
\mbox{}\ \ \ \ \ \ \ \ \ \ \{\textsf{t}:=\textsf{alloc}(\textsf{Node}); \ \ \textsf{t}\rightarrow k := \textsf{k};\ \ \textsf{t}\rightarrow d:=\textsf{d}\}\\
\outlying(\rho,\texttt{Block}(\textsf{t}))\land(\textsf{t}\neq \textbf{nil})\land(\textsf{t}\rightarrow l = \textbf{nil})\land ( \textsf{t}\rightarrow r=\textbf{nil})\land \\
(\textsf{t}\rightarrow k = \textsf{k})\land(\textsf{t}\rightarrow d=\textsf{d})
\end{array}$$

\end{example}

\subsection{The proof rule for \textbf{if}-statement}
The proof rule for if-statement is almost same as the one in Hoare's logic.
The only difference is that the conditional expression should always evaluate to a boolean value.
$$\frac{\ \ \ \ (p \land e)\{ s_1\}q \ \ \ \ (p\land \neg e)\{s_2\} q\ \ \ }
{\ \ \ p\land (e\lor \neg e)\ \{\mbox{ \texttt{if} }(e)\ s_1\mbox{
\texttt{else} }s_2\ \}\ q\ \ \ }\ \ \ \ \ \ \ \ \ \ (\textrm{IF-ST})$$

\subsection{The proof rule for \textbf{while}-statement}
The proof rule in our logic is essentially same as the corresponding one in Hoare's logic.
The only difference is that we require that the invariant implies the definedness of the conditional expression.
We have the following rule.

$$
\frac{\ \ \ \ (p \land e)\{s\}p\land (e \lor \neg e)\ \ \ }
{\ \ \ p\land(e\lor \neg e)\ \{\mbox{ \texttt{while} }(e)\ \ s\ \} \neg e \land p\ \ \ }
 \ \ \ \ \ (\textrm{WHILE-ST})$$
\begin{center}where $p$ is a formula containing no pre-state term.\end{center}

\subsection{The proof rule for sequential statements}
The two statements in a sequential composition $s_1;s_2$ execute sequentially. If we prove the assertions about $s_1$ and $s_2$ respectively,
we can compose these two assertions into an assertion about $s_1;s_2$. As the pre-state of $s_1$ is different from that of $s_2$,
pre-state terms in two assertions should be eliminated.
\begin{center}
$$\frac{\ \ \ p\,\{s_1\}\,q \ \ \ \ \ q\,\{s_2\}\,r\ \ \ }
{p\,\{s_1;s_2\}\,r}\ \ \ \ \ \ \ \ \ \ (\textrm{SEQ-ST})$$
where $p$, $q$ and $r$ contain no pre-state term.
\end{center}

\begin{example}
In the proof rules for while-statements and sequential statements, the pre-/post-conditions contain no pre-state terms.
To prove the relation between pre-state and post-state, we can first prove an assertion with free logical variables,
and then put pre-state terms into it using the proof rule \textrm{SUBSTITUTION}, which will be presented in
Section~\ref{SEC-OTHER-RULES}.

Let $\textsf{x}, \textsf{y}, \textsf{z}$ be three integer program variables. $v_x,v_y,v_z$ be three logic variables.  We have the following two assertions.
$$\begin{array}{c}
\outlying(\rho_1, \{\&\textsf{x}\})\ \{\textsf{x}:=\textsf{x}+1;\}\  \rho_1\land \textsf{x}=\overleftarrow{\textsf{x}}+1\\
\outlying(\rho_2, \{\&\textsf{y}\})\ \{\textsf{y}:=\textsf{y}-\textsf{z};\}\  \rho_2\land \textsf{y}=\overleftarrow{\textsf{y}}-\textsf{z}\\
\end{array}$$
Substitute $\rho_1$ and $\rho_2$ with $\outlying(\rho,\{\&\textsf{x},\&\textsf{y}\})\land(v_x=\overleftarrow{\textsf{x}})\land (v_y={\textsf{y}})$ and
$\outlying(\rho,\{\&\textsf{x},\&\textsf{y}\})\land({\textsf{x}}=v_x+1)\land (v_y=\overleftarrow{\textsf{y}})$ respectively, then apply the consequence rule (presented later),
we have
$$\begin{array}{l}
\outlying(\rho, \{\&\textsf{x},\&\textsf{y}\})\land(v_x={\textsf{x}})\land (v_y={\textsf{y}})\\
\mbox{}\ \ \ \ \ \{\textsf{x}:=\textsf{x}+1;\}\\
\outlying(\rho, \{\&\textsf{x},\&\textsf{y}\})\land(v_y={\textsf{y}})\land (\textsf{x}=v_x+1)\\
\\
\outlying(\rho, \{\&\textsf{x},\&\textsf{y}\})\land({\textsf{x}}=v_x+1)\land(v_y={\textsf{y}})\\
\mbox{}\ \ \ \ \ \{\textsf{y}:=\textsf{y}-\textsf{z};\}\\
\outlying(\rho, \{\&\textsf{x},\&\textsf{y}\})\land ({\textsf{x}}=v_x+1) \land(\textsf{y}=v_y-\textsf{z})
\end{array}$$
Apply the proof rule for sequential statements, we have
$$\begin{array}{l}
\outlying(\rho, \{\&\textsf{x},\&\textsf{y}\})\land(v_x={\textsf{x}})\land (v_y={\textsf{y}})\\
\mbox{}\ \ \ \ \ \{\textsf{x}:=\textsf{x}+1;\ \ \ \textsf{y}:=\textsf{y}-\textsf{z};\}\\
\outlying(\rho, \{\&\textsf{x},\&\textsf{y}\})\land ({\textsf{x}}=v_x+1) \land(\textsf{y}=v_y-\textsf{z})
\end{array}$$
Applying the rule $\textrm{SUBSTITUTION}$, substitute $v_x$ and $v_y$ with $\overleftarrow{\textsf{x}}$ and $\overleftarrow{\textsf{y}}$ respectively, we have
$$\begin{array}{l}
\outlying(\rho, \{\&\textsf{x},\&\textsf{y}\})\\
\mbox{}\ \ \ \ \ \{\textsf{x}:=\textsf{x}+1;\ \ \ \textsf{y}:=\textsf{y}-\textsf{z};\}\\
\outlying(\rho, \{\&\textsf{x},\&\textsf{y}\})\land ({\textsf{x}}=\overleftarrow{\textsf{x}}+1) \land(\textsf{y}=\overleftarrow{\textsf{y}}-\textsf{z})
\end{array}$$
Substitute $\rho$ with $\outlying(\rho_0,\{\&\textsf{x},\&\textsf{y}\})\land (v'_y=\textsf{z}*\overleftarrow{\textsf{x}}+\overleftarrow{\textsf{y}})$, and applying the consequence rule, we have
$$\begin{array}{l}
\outlying(\rho_0, \{\&\textsf{x},\&\textsf{y}\})\land(v_y'=\textsf{z}*{\textsf{x}}+{\textsf{y}})\\
\mbox{}\ \ \ \ \ \{\textsf{x}:=\textsf{x}+1;\ \ \ \textsf{y}:=\textsf{y}-\textsf{z};\}\\
\outlying(\rho_0, \{\&\textsf{x},\&\textsf{y}\})\land (v_y'=\textsf{z}*{\textsf{x}}+{\textsf{y}})
\end{array}$$
Apply the proof rule for while-statements, we have
$$\begin{array}{l}
\outlying(\rho_0, \{\&\textsf{x},\&\textsf{y}\})\land(v_y'=\textsf{z}*{\textsf{x}}+{\textsf{y}})\\
\mbox{}\ \ \ \ \ \{\textbf{while }(\textsf{y}\ge \textsf{z})\textbf{ begin }\textsf{x}:=\textsf{x}+1;\ \ \ \textsf{y}:=\textsf{y}-\textsf{z};\textbf{ end }\}\\
\outlying(\rho_0, \{\&\textsf{x},\&\textsf{y}\})\land (v_y'=\textsf{z}*{\textsf{x}}+{\textsf{y}})\land (\textsf{y}<\textsf{z})
\end{array}$$
From the proof rule $\textrm{SUBSTITUTION}$, we have
$$\begin{array}{l}
\outlying(\rho_0, \{\&\textsf{x},\&\textsf{y}\})\land(\overleftarrow{\textsf{y}}=\textsf{z}*{\textsf{x}}+{\textsf{y}})\\
\mbox{}\ \ \ \ \ \{\textbf{while }(\textsf{y}\ge \textsf{z})\textbf{ begin }\textsf{x}:=\textsf{x}+1;\ \ \ \textsf{y}:=\textsf{y}-\textsf{z};\textbf{ end }\}\\
\outlying(\rho_0, \{\&\textsf{x},\&\textsf{y}\})\land (\overleftarrow{\textsf{y}}=\textsf{z}*{\textsf{x}}+{\textsf{y}})\land (\textsf{y}<\textsf{z})
\end{array}$$
\end{example}

%
%

\section{Other proof rules}\label{SEC-OTHER-RULES}
In this section, we presents the rest of the proof rules in our logic. Most of them are adopted directly from Hoare's Logic.
The consequence rule says that provided an assertion, we can strengthen the pre-condition, or weaken the post-condition.
$$
\frac{\ \ \ p\,\{s\}\,q \ \ \ \ p'\Rightarrow p\ \ \ \ \ q\Rightarrow q'\ \ \ }
{p'\,\{s\}\,q'}\ \ \ \ \ \ \ \ \ \ \ (\textrm{CONSEQ})
$$
The proof rules \textrm{CONJUNCTION} and \textrm{DISJUNCTION} is same as the ones in Hoare's logic.
$$\frac{\ \ \ p\,\{s\}\,q \ \ \ \ p'\,\{s\}\,q'\ \ \ }
{p\land p'\,\{s\}\,q\land q'}\ \ \ \ \ \ \ \ \ \ (\textrm{CONJUNCTION})
$$
$$
\frac{\ \ \ p\,\{s\}\,q \ \ \ \ p'\,\{s\}\,q'\ \ \ }
{p\lor p'\,\{s\}\,q\lor q'}\ \ \ \ \ \ \ \ \ \ (\textrm{DISJUNCTION})
$$
The proof rules \textrm{ALL} and \textrm{EXIST} are respectively the generalized version of \textrm{CONJUNCTION} and \textrm{DISJUNCTION}.
$$
\frac{\ \ \ p\,\{s\}\,q\ \ \ }{\forall x\cdot p\,\{s\}\forall x\cdot q}\ \ \mbox{where $x$ is arbitrary}\ \ \ \ \ \ \ \ \ \ (\textrm{ALL})
$$
$$
\frac{\ \ \ p\,\{s\ \}\,q\ \ \ }{\exists x\cdot p\,\{s\}\exists x\cdot q}\ \ \mbox{where $x$ is arbitrary}\ \ \ \ \ \ \ \ \ \ (\textrm{EXIST})
$$
The proof rule \textrm{SUBSTITUTION} says that we can substitute a variable with a term $e$ and its pre-state form respectively in the pre-condition and the post-condition, as long as $e$ is denoting in the pre-state.
$$\frac{\ \ \ p\,\{s\}\,q\ \ \ p\Rightarrow e\downarrow \ \ \ \ }{p[e/x]\,\{s\} q[\overleftarrow{e}/x]}\ \ \mbox{where $x$ is arbitrary}\ \ \ \ \ \ \ \ \ \ (\textrm{SUBSTITUTION})$$

\section{Concise form of proofs}\label{SEC-CONCISE-FORM}
For conciseness, we write code in the following manner.
\begin{itemize}
\item For a sequential composition of $s_1;s_2$, we write
$$\begin{array}{c}
\{p \}\\
s_1\\
\{q\}\\
s_2\\
\{r\}
\end{array}$$ to show that we can first prove $p \{s_1\} q$ and $q\{s_2\}r$, then derive the assertion $p\{s_1,s_2\} r$ by the sequential statement rule.
It is required that $p,q,r$ contain no pre-state term.
\item For a if-statement $\textbf{if }(e)\ s_1\textbf{ else }s_2$, we write
   $$\begin{array}{l}
   \{p \land (e\lor \neg e)\}\\
   \textbf{if }(e)\\
   \mbox{}\ \ \ \{p\land e\}\\
   \mbox{}\ \ \ s_1\\
   \mbox{}\ \ \ \{q\}\\
   \textbf{ else }\\
   \mbox{}\ \ \ \{p\land \neg e\}\\
   \mbox{}\ \ \ s_2\\
   \mbox{}\ \ \ \{q\}\\
   \{q\}
\end{array}$$
to show that we can first prove $p\land e\ \{s_1\}\ q$ and $p\land \neg e\ \{s_2\} q$, then derive
$p\land (e\lor \neg e)\ \{\textbf{if }(e)\ s_1\textbf{ else }s_2\}\ q$ by the proof rule for if-statements.
\item For a while-statement $\textbf{while }(e)\textbf{ do }s$, we
write
   $$\begin{array}{l}
   \{p \land (e\lor \neg e) \}\\
   \textbf{while }(e)\\
   \mbox{}\ \ \ \{p\land e\}\\
   \mbox{}\ \ \ s\\
   \mbox{}\ \ \ \{p\land (e\lor \neg e)\}\\
   \{p\land \neg e\}
\end{array}$$
to show that we can first prove that $p$ is the invariant of this while-statement, then get the assertion $p\land (e\lor \neg e)\ \{\textbf{while }(e)\textbf{ do }s\}\ p\land \neg e$.
\item An application of the consequence rule
$$
\frac{\ \ \ p\,\{s\}\,q \ \ \ \ p'\Rightarrow p\ \ \ \ \ q\Rightarrow q'\ \ \ }
{p'\,\{s\}\,q'}
$$
is written as
$$\begin{array}{rl}
&\{p'\}\\
\Rightarrow &\{p\}\\
&s\\
&\{q\}\\
\Rightarrow &\{q'\}
\end{array}
$$
\item An substitution of predicate variables or an application of Theorem~\ref{}
$$\begin{array}{lll}
\{p'\}                              &\mbox{}\ \ \ \ \ \ \ \ \ \ \ \ \  & \{p\land \overline{r}\}\\
\ \ \uparrow x \rightsquigarrow e   &                       &\ \ \uparrow \rho \rightsquigarrow \overline{r}       \\
\{p\}                               &                       &\{\mathtt{Outlying}(\rho,e)\land p\}\\
s                                   &                       &s\\
\{q\}                               &                       &\{\rho\land q\}\\
\ \ \downarrow x \rightsquigarrow \overleftarrow{e} &       &\ \ \downarrow \rho \rightsquigarrow r\\
\{q'\}                              &                       &\{q\land r\}
\end{array}
$$
Please be noticed that according to the proof rule SUBSTITUTION, if we substitute $x$ with $e$, $q'$ is derived by substitute $x$ in $q$ with $\overleftarrow e$.
When substitute $\rho$ with $r$, we must prove the second premise, i.e. $p \land\overline{r}\Rightarrow (\overline{\mathfrak{M}(r)}\cap e=\emptyset)$.
\end{itemize}

\section{The proof of the running example in the concise form}\label{SEC-PROOF}
Let $MSet$ be $$\{\&\textsf{pt}\}\cup \textsf{FldD}(\textsf{root})$$
Let $\textrm{INV}$ be the formula
$$\begin{array}{l}
\outlying(\rho_0,MSet)\land\textsf{isHBST}({\textsf{root}})\land \textsf{Map}(\textsf{root})=x\land \\ ({\textsf{pt}\neq \textbf{nil}}) \land (\textsf{pt}\in \textsf{NodeSet}(\textsf{root}))\land (\textsf{k}\in \textsf{Dom}(\textsf{pt}))
\end{array}
$$

The proof of the specification \ref{WHOLE-SPECIFICATION} in the concise form is depicted in Figure~\ref{EXAMPLE-W-PROOF}.
The following are the properties about the recursive functions used in this proof.
$$\begin{array}{l}
\textsf{isHBST}(x)\Rightarrow
\&\textsf{pt}\not\in \textsf{MSF}_{lr}(x)\cup\textsf{MSF}_{lrk}({x})\cup \textsf{MSF}_{lrkd}({x})\\
\\
\textsf{isHBST}(x)\land \textsf{pt}\in \textsf{NodeSet}(x)\Rightarrow\\
\mbox{\ \ \ \ \ \ \ \ \ }\&\textsf{pt}\rightarrow D \not\in \textsf{MSF}_{lrk}(x)\cup\textsf{MSF}_{lr}({x})\cup\textsf{MPP}_m({x},\textsf{pt})\\
\\
\textsf{isHBST}(x)\land ( y\in \textsf{Dom}(x) ) \land (y<x\rightarrow K)\Rightarrow
y\in\textsf{Dom}(x\rightarrow l)\\
\\
\textsf{isHBST}(x)\land(y\in \textsf{Dom}(x))\land (y>x\rightarrow K) \Rightarrow
y\in\textsf{Dom}(x\rightarrow r)\\
\\
\textsf{isHBST}(x)\land(y \in \textsf{NodeSet}(x))\Rightarrow\\
\mbox{\ \ \ \ \ \ \ \ \ }\textsf{Map}(x)=\textsf{MapP}(x,y)\dag \{y\rightarrow K\mapsto y\rightarrow D\}\\
\\
\textsf{NodeSet}(x):\textbf{SetOf}(\textbf{Ptr})\Rightarrow x\in \textsf{NodeSet}(x)
\end{array}$$

\begin{figure}
\begin{center}
\parbox{300pt}{
\begin{tabbing}
\mbox{}\ \ \ \ \=\ \ \ \ \=\ \ \ \ \ \ \=\ \ \ \ \ \ \=\\
    \>\{$\texttt{Outlying}(\rho_0,MSet)\land \textsf{isHBST}(\textsf{root})\land$ \\
    \>\ \ $\textsf{Map}(\textsf{root})=\overleftarrow{\textsf{Map}(\textsf{root})} \land\textsf{k} \in \textsf{Dom}(\textsf{root})$\}\\
    \>  \>$\uparrow x\rightsquigarrow {\textsf{Map}(\textsf{root})}$\\
    \>\{$\texttt{Outlying}(\rho_0,MSet)\land \textsf{isHBST}(\textsf{root})\land$ \\
    \>\ \ $\textsf{Map}(\textsf{root})=x \land\textsf{k} \in \textsf{Dom}(\textsf{root})$\}\\
    \>$\textsf{pt}:=\textsf{root};$\\
    \>\{$\textrm{INV}$\}\\
    \>\textbf{while} ($\textsf{pt}\!\rightarrow\!\!K \neq \textsf{k}$)\\
    \>\textbf{begin}\\
    \>  \>\{$\textrm{INV} \land (\textsf{pt}\!\rightarrow\!\!K \neq \textsf{k}) $\}\\
    \>  \>\textbf{if} ($\textsf{k} < \textsf{pt}\rightarrow K$ )\\
    \>  \>  \>\{$\textrm{INV} \land (\textsf{pt}\!\rightarrow\!\!K \neq \textsf{k})\land (\textsf{k} < \textsf{pt}\rightarrow K ) $\}\\
    \>  \>\ \ $\Rightarrow$\>$\{\outlying(\rho_0,MSet)\land\textsf{isHBST}({\textsf{root}})\land \textsf{Map}(\textsf{root})=x\land$\\
    \>  \>  \>\ \ $({\textsf{pt}\rightarrow l\neq \textbf{nil}}) \land (\textsf{pt}\rightarrow l\in \textsf{NodeSet}(\textsf{root}))\land (\textsf{k}\in\textsf{Dom}(\textsf{pt}\rightarrow l))\}$\\
    \>  \>  \>$\textsf{pt} := \textsf{pt}\rightarrow l$\\
    \>  \>  \>$\{\outlying(\rho_0,MSet)\land\textsf{isHBST}({\textsf{root}})\land \textsf{Map}(\textsf{root})=x\land$\\
    \>  \>  \>\ \ $({\textsf{pt}\neq \textbf{nil}}) \land (\textsf{pt}\in \textsf{NodeSet}(\textsf{root}))\land (\textsf{k}\in \textsf{Dom}(\textsf{pt}))\}$\\
    \>  \>\ \ $\Rightarrow$\>\{$\textrm{INV}$\}\\
    \>  \>\textbf{else}\\
    \>  \>  \>\{$\textrm{INV} \land (\textsf{pt}\!\rightarrow\!\!K \neq \textsf{k})\land\neg (\textsf{k} < \textsf{pt}\rightarrow K ) $\}\\
    \>  \>\ \ $\Rightarrow$\>$\{\outlying(\rho_0,MSet)\land\textsf{isHBST}({\textsf{root}})\land \textsf{Map}(\textsf{root})=x\land$\\
    \>  \>  \>\ \ $({\textsf{pt}\rightarrow r\neq \textbf{nil}}) \land (\textsf{pt}\rightarrow r\in \textsf{NodeSet}(\textsf{root}))\land (\textsf{k}\in\textsf{Dom}(\textsf{pt}\rightarrow r))\}$\\
    \>  \>  \>$\textsf{pt} := \textsf{pt}\rightarrow r$;\\
    \>  \>  \>$\{\outlying(\rho_0,MSet)\land\textsf{isHBST}({\textsf{root}})\land \textsf{Map}(\textsf{root})=x\land$\\
    \>  \>  \>\ \ $({\textsf{pt}\neq \textbf{nil}}) \land (\textsf{pt}\in \textsf{NodeSet}(\textsf{root}))\land (\textsf{k}\in \textsf{Dom}(\textsf{pt}))\}$\\
    \>  \>\ \ $\Rightarrow$\>\{$\textrm{INV}$\}\\
    \>  \>\{$\textrm{INV}$\}\\
    \>\textbf{end}\\
    \>\{$\textrm{INV}\land \neg(\textsf{pt}\!\rightarrow\!\!K \neq \textsf{k})$\}\\
$\Rightarrow$\> $\{\outlying(\rho_0,MSet)\land\textsf{isHBST}({\textsf{root}})\land$\\
    \>\ \ $\textsf{MapP}(\textsf{root}, \textsf{pt})\dag\{\textsf{pt}\rightarrow K\mapsto \textsf{d}\}=x\dag\{ \textsf{k}\mapsto \textsf{d}\}\land$ \\
    \>\ \ $({\textsf{pt}\neq \textbf{nil}}) \land (\textsf{pt}\in \textsf{NodeSet}(\textsf{root}))\land (\textsf{k}\in \textsf{Dom}(\textsf{pt}))\}$\\
    \>  \>  \>$\uparrow \rho_1  \rightsquigarrow \outlying(\rho_0,MSet)\land\textsf{isHBST}({\textsf{root}})\land\dots \land (\textsf{k}\in \textsf{Dom}(\textsf{pt}))$\\
    \>$\{\outlying(\rho_1,\&\textsf{pt}\rightarrow D)\land \textsf{pt}\neq \textbf{nil} \}$\\
    \>$\textsf{pt}\rightarrow D := \textsf{d}$;\\
    \>$\{\rho_1\land \textsf{pt}\rightarrow D = \textsf{d}\}$\\
    \>  \>  \>$\downarrow \rho_1  \rightsquigarrow \outlying(\rho_0,MSet)\land\textsf{isHBST}({\textsf{root}})\land\dots \land (\textsf{k}\in \textsf{Dom}(\textsf{pt}))$\\
    \>$\{\outlying(\rho_0,MSet)\land\textsf{isHBST}({\textsf{root}})\land$\\
   \>\ \ $\textsf{MapP}(\textsf{root},\textsf{pt})\dag\{\textsf{pt}\rightarrow K\mapsto \textsf{pt}\rightarrow D\}=x\dag\{ \textsf{k}\mapsto \textsf{d}\}\land$ \\
    \>\ \ $({\textsf{pt}\neq \textbf{nil}}) \land (\textsf{pt}\in \textsf{NodeSet}(\textsf{root}))\land (\textsf{k}\in \textsf{Dom}(\textsf{pt}))\}$\\
$\Rightarrow$\>$\{\rho_0\land \textsf{isHBST}(\textsf{root})\land \textsf{Map}(\textsf{root})=x\dag\{\textsf{k}\mapsto \textsf{d}\}$\}\\
    \>  \> $\downarrow x\rightsquigarrow\overleftarrow{\textsf{Map}(\textsf{root})}$\\
    \>$\{\rho_0\land \textsf{isHBST}(\textsf{root})\land \textsf{Map}(\textsf{root})=\overleftarrow{\textsf{Map}(\textsf{root})}\dag\{\textsf{k}\mapsto \textsf{d}\}$\}
\end{tabbing}}
\end{center}
\caption{The proof in code of the running example}\label{EXAMPLE-W-PROOF}
\end{figure}

\section{Conclusions and further works}\label{SEC-CONCLUSION}
In this paper, we present an extension of Hoare logic to deal with pointer programs. The pre-conditions and post-conditions are logic formulas with recursive functions.

Three functions ($\ast$, $\&\rightarrow n$ and $\&[\,]$) are respectively used to represents memory unit access,
record-field access, and array-element access. These functions are specified by a set of axiom. These axioms formally
specify our memory models with the following properties.
\begin{itemize}
\item A non-nil pointer always refers to a valid memory unit or memory block.
\item Two different memory blocks are either disjoint, or one of them contains another one.
\item Each declared program variable is assigned an unique memory block.
\item A memory block with a composite type is always allocated as a whole. The memory blocks
allocated for different components of a composite type are disjoint with each other.
\end{itemize}

To support local reasoning, we use specifications of the form
$$\rho\land (\mathfrak{M}(\rho)\cap e =\emptyset )\land pre\ \ \{\ s\ \}\ \ \rho\land post$$
In this specification, $e$ represents an upper bound of the memory set modified by $s$. We can
substitute $\rho$ with proper predicates in this specification to get global specifications.

We use \emph{pre-state terms} to denoting values before program executions. Pre-state terms can
appear in both pre-conditions and post-conditions. A pre-state term $\overleftarrow e$ denoting
the value of $e$ interpreted based on the pre-state. In pre-conditions,
$\overleftarrow e$ and $e$ is interchangeable. The difference between $\overleftarrow e$ and $e$ is that
the memory scope form of $\overleftarrow e$ is $\emptyset$.

On a specific program state, the value of a term $e$ relies only on a finite set of memory units.
This value of $e$ does not change if the contents stored in these memory units keep unmodified. We show that such a 
memory unit set can be expressed using a term constructed syntactically. Given a user-defined recursive function,
we can construct another recursive function to represents the set of memory units accessed by this
function during its evaluation process. We use an axiom to specify the properties about memory scope
terms and memory scope functions.

For the program statements, we present an axiom for assignment statements, and an axiom for memory
allocation statements. Other axioms and proof rules are same as the ones in Hoare's logic.

This logic has the following advantages.
\begin{itemize}
\item The axioms and proof rules in our logic is simpler comparing to those of Separation Logic. In Separation Logic, to deal with
    the separation conjunction, 14 axioms are introduced for general cases. Besides these, five special classes of formulas are defined based on the
    semantics of the separation conjunction. More axioms and theorems are given for these special formulas. To apply these
    special axioms and rules, people have to judge wether the formula is really in the class based on the semantics of separation conjunction.  
\item This logic deals with high-level program types (record/array) directly. Separation logic
    introduces two new logical symbols ($\stackrel{[e]}{\hookrightarrow}$, $ \bigodot$) and 11 axioms to deal with records and arrays.
    People still have to deal with composite types with low-level address manipulation.
\item This logic is easy to learn. Most of the knowledge (for example, recursive functions, FOL, memory layout
    of composite P-types.) in this logic have been (explicitly or implicitly) taught in undergraduate CS courses.
    For examples, the concept of recursive functions and first order logic are already taught in undergraduate CS
    courses. The proof rules about program variables, $\ast$, $\&\rightarrow n$, $\&[\,]$ are taught informally
    in the undergraduate courses about programming languages and compilers.
\item This logic supports reuse of proofs. Most of recursive functions and their proved properties are about data
    structures. They are independent of the code under verification. So these properties can be reused in verification
    of other code using same data structures. It is possible to build a library of pre-defined recursive functions,
    their memory-scope functions, and their verified properties.
\item People can easily specify and verify relations between pre-/post-states with pre-state terms introduced in
    our logic.
\item Our logic supports local reasoning precisely. The upper-bound of memory units modified by the program can be specified by a term.   
\end{itemize}

We also use several examples to show the efficacy of our logic. The example in the Appendix~\ref{SEC-SEL-SORT} shows that our logic can also
deal with program templates, i.e. a program with some  components indirectly specified.

In the future, we will extend our logic to deal with more programming language concepts: function calls, function pointers, class/object, generics, and so on.
At the mean time, we will try to build a library of pre-defined recursive functions, their memory scopes, and their properties for frequently used data structures.

\appendix

\section{Another example: the Schorre-Waite algorithm}\label{SEC-SHORRE-WAITE}
\subsection{The Program}
\subsubsection{Types and Variables}
The Schorre-Waite program to be verified, together with a type definition and three variable declarations, is depicted in Figure~\ref{PROGRAM}.
In the program, we use some abbreviations for conciseness.
\begin{center}
\begin{tabular}{|c|c|}
\hline
$\textsf{CONDITION}_{wh}$ &$(\textsf{p}\neq \textbf{nil}\, ?\, \textbf{true} : ((\textsf{t} = \textbf{nil})\, ?\, \textbf{false} : \neg \textsf{t}\rightarrow\textsf{mk})$\\
\hline
$\textsf{CONDITION}_1$ & $(\textsf{t} = \textbf{nil}\,?\, \textbf{true} :\textsf{t}\rightarrow\textsf{mk})$\\
\hline
$\textsf{CONDITION}_{2}$  &  $(\textsf{p}\rightarrow\textsf{chk})$\\
\hline
\end{tabular}
\end{center}
The program is an elaborative implementation of the depth first search algorithm.
The search path is stored in a linked list, of which the links reuses the memory cells for the field $\textsf{l}$ or $\textsf{r}$.

\begin{figure}
\begin{tabbing}
\ \ \ \ \ \ \=\ \ \ \ \=\ \ \ \ \ \=\ \ \ \ \ \=\ \ \ \ \ \=\ \ \ \ \ \=\\
Type $T$: $\textbf{Rec}(\textsf{l}: \textbf{P}(\textsf{T}),\textsf{r}: \textbf{P}(\textsf{T}), \textsf{chk}: \textbf{boolean}, \textsf{mk}: \textbf{boolean}))$\\
Three variables: $\textsf{root}, \textsf{t}, \textsf{p}$ declared with type $\textbf{P}(\textsf{T})$.\\
\\
(1) \>$\textsf{t} := \textsf{root};$\\
(2) \>$\textsf{p} := \textbf{nil};$\\
(3) \>$\textbf{while }( \textsf{CONDITION}_{wh}  )\textbf{ do } \mbox{//stack not empty or \textsf{t} not explored;}$\\
(4) \>$\textbf{begin}$\\
(5)	\>  \>$\textbf{if }(\textsf{CONDITION}_{1}) \mbox{//t is already explored or t is \textbf{nil};}$\\
(6)	\>  \>$\textbf{then}$\\
(7) \>  \>  \>$\textbf{ if }$\=$(\textsf{CONDITION}_2)\textbf{ then}$  //POP;\\
(8)	\>  \>  \>  \>\textbf{begin }$\textsf{q}:=\textsf{t};$\ \ \ $\textsf{t}:=\textsf{p};$\ \ \ $\textsf{p} := \textsf{p}\rightarrow\textsf{r};$\ \ $\textsf{t}\rightarrow\textsf{r} := \textsf{q};$\textbf{ end}\\
(9)	\>  \>  \>\textbf{else} //TRY right neighbor;\\
(10)	\>  \>  \>  \>\textbf{begin }\\
(11)\>  \>  \>  \>  \>$\textsf{q}:=\textsf{t}$;\ \ \ $\textsf{t}:=\textsf{p}\rightarrow\textsf{r};$\ \ $\textsf{p}\rightarrow\textsf{r} := \textsf{p}\rightarrow\textsf{l};$\\
(12)\>  \>  \>  \>  \>$\textsf{p}\rightarrow\textsf{l} := \textsf{q}; $\ \ \ $\textsf{p}\rightarrow\textsf{chk} := \textbf{true};$\\
(13)\>  \>  \>  \>\textbf{end}\\
(14)\>  \>\textbf{else}	//$\textsf{t}$ is not explored.  PUSH $\textsf{t}$ into the stack;\\
(15)\>	\>   \>\textbf{begin }\\
(16)\>  \>  \>  \>$\textsf{q} := \textsf{p};$ \ $\textsf{p} := \textsf{t};$ \ $\textsf{t} := \textsf{t}\rightarrow\textsf{l};$\ \ $\textsf{p}\rightarrow\textsf{l} := \textsf{q};$\\
(17)\>  \>  \>  \>$\textsf{p}\rightarrow\textsf{mk} := \textbf{true};$\ \ $\textsf{p}\rightarrow\textsf{chk} := \textbf{false};$\\
(18)\>  \>  \>\textbf{end}\\
(19)\>\textbf{end}\\
\end{tabbing}
\caption{The Schorr-Waite Algorithm}\label{PROGRAM}
\end{figure}

\subsection{Recursive functions}\label{SW-FUNC}
We use two constants $\texttt{L}_0$ and $\texttt{R}_0$to represent the edges. They are two maps. For a node $x$ in the graph, $\texttt{L}_0(x)$ and $\texttt{R}_0(x)$ are respectively the left-neighbor and right-neighbor of $x$ in the initial graph. The constants $\texttt{r}_0$ is the starting node. The constant $\texttt{NS}_0$ represents the smallest set satisfying the following properties.
\begin{itemize}
\item $\texttt{r}_0\in \texttt{NS}_0$;
\item $x\in \texttt{NS}_0 \land \texttt{L}_0(x)\neq \textbf{nil} \Rightarrow \texttt{L}_0(x)\in \texttt{NS}_0$;
\item $x\in \texttt{NS}_0 \land \texttt{R}_0(x)\neq \textbf{nil} \Rightarrow \texttt{R}_0(x)\in \texttt{NS}_0$.
\end{itemize}

The definitions of the recursive functions used in the specification and the proof are depicted in Figure~\ref{SW-FUNC}.
The function $\textsf{Stack(x)}$ retrieves the segment of the current path starting from $x$.
The boolean-typed function $\textsf{isAcyclic}(x)$ asserts that the list $x$ is acyclic.
The boolean-typed function $\textsf{PathProp}$ asserts the properties about the current path.

The memory scope over-approximations for some functions are as follow.
\begin{center}
\begin{tabular}{|c|c|}
\hline
Recursive Functions     &   Memory scope over-approximations\\
\hline
$\textsf{NextNode}(x)$  &   $(x=\textbf{nil})?\emptyset: \&x\rightarrow \textsf{chk}$\\
\hline
$\textsf{Explored}(x)$    &   $(x=\textbf{nil}) ? \emptyset : \{\& x\rightarrow \textsf{mk}\}$\\
\hline
$\textsf{Stack}(x)$       &   $\{\texttt{Block}(x) | x\in \textsf{rng}(\textsf{Stack}(x))\}$\\
\hline
$\textsf{isAcyclic}(s)$   &   $\emptyset$\\
\hline
\end{tabular}
\end{center}
Let $\textsf{MarkedFldMk}$ be $\{\&y\rightarrow\textsf{mk}|y\in\texttt{NS}_0\land y\rightarrow\textsf{mk}\}$.
As $$\textsf{Explored}(x)\land x\in \texttt{NS}_0\Rightarrow \& x\rightarrow \textsf{mk}\in \textsf{MarkedFldMk}$$ we have the following properties.
$$
\begin{array}{l}
\textsf{InPath}(x)\land \textsf{rng}(\textsf{Stack}(x))\subseteq \texttt{NS}_0\Rightarrow\\
 \mbox{}\ \ \ \ \ \ \ \mathfrak{M}(\textsf{InPath})(x)\subseteq \{\texttt{Block}(y) | y\in \textsf{rng}(\textsf{Stack}(x))\}\cup \textsf{MarkedFldMk}\\
 \\
\textsf{OutPath}(s)\land s\subseteq \texttt{NS}_0\Rightarrow\\
\mbox{}\ \ \ \ \ \ \mathfrak{M}(\textsf{OutPath})(s)\subseteq \{\texttt{Block}(x)|x\in s\}\cup \textsf{MarkedFldMk}
\end{array}
$$
\begin{figure}
$$\begin{array}{l}
\textsf{NextNode}(x:\textbf{P}(T)):\textbf{P}(T))\triangleq (x=\textbf{nil})? \texttt{r}_0 : (x\rightarrow \textsf{chk} ? \texttt{R}_0(x) : \texttt{L}_0(x))\\
\textsf{Explored}(x:\textbf{P}(T)):\textbf{boolean} \triangleq (x=\textbf{nil}) ? \textbf{true} : x\rightarrow \textsf{mk}\\
\textsf{Stack}(x:\textbf{P}(T)):\textsf{SeqOf}(\textbf{P}(T))\triangleq\\
\mbox{}\ \ \ \ \ \ \ \ (x=\textbf{nil}) ? [\,] : x^\frown \textsf{Stack}(x\rightarrow \textsf{chk} ? x\rightarrow \textsf{r} : x\rightarrow \textsf{l})\\
\\
\textsf{isAcyclic}(s:\textbf{SeqOf}(\textbf{P}(T))):\textbf{boolean} \triangleq \\
\mbox{\ \ \ \ \ \ \ \ \ }s= []? \textbf{true} : \textsf{head}(s)\not\in \textsf{rng}(\textsf{tail}(s)) \land \textsf{isAcyclic}(\textsf{tail}(s))\\
\\
\textsf{InPath}(x:\textbf{P}(T)):\textbf{boolean} \triangleq x=\textbf{nil} ? \textbf{true} :\\
\mbox{}\ \ \ \ \ \ x\in \texttt{NS}_0\land x\rightarrow \textsf{mk} = \textbf{true} \land\\
\mbox{}\ \ \ \ \ \left(\begin{array}{l}
x\rightarrow \textsf{chk} ? ((\textsf{NextNode}(x\rightarrow \textsf{r})= x) \land (x\rightarrow \textsf{l}=\texttt{L}_0(x))\land \textsf{Explored}(\texttt{L}_0(x)) :\\
\mbox{}\ \ \ \ \ \ \ \ \ ((\textsf{NextNode}(x\rightarrow\textsf{l})=x) \land (x\rightarrow \textsf{r} = \texttt{R}_0(x))\\
\end{array}\right)\land\\
\mbox{}\ \ \ \ \ \ \textsf{InPath}(x\rightarrow \textsf{chk} ? x\rightarrow \textsf{r}: x\rightarrow \textsf{l})\\
\\
\textsf{OutPath}(x:\textbf{SetOf}(\textbf{P}(T)):\textbf{boolean} \triangleq\\
\mbox{}\ \ \ \ \ \ \bigwedge_{y\in x}
\left(\begin{array}{l}
x\rightarrow \textsf{l}=\texttt{L}_0(x) \land x\rightarrow \textsf{r} = \texttt{R}_0(x) \land\\
(x\rightarrow \textsf{mk} ? (x\rightarrow\textsf{chk}\land\textsf{Explored}(\texttt{L}_0(x))\land\textsf{Explored}(\texttt{R}_0(x))): \textbf{true} )
\end{array}\right)
\\
\end{array}$$\\
\caption{The recursive functions used in the proof}\label{SW-FUNC}
\end{figure}

\subsection{The specification}
Let $MSet$ be $\{\&\textsf{p},\&\textsf{t},\&\textsf{q}\}\cup\{\texttt{Block}(x)|x\in \texttt{NS}_0\}$.
Notice that $\mathfrak{M}(Set)$ is $\emptyset$. The program can be specified as follow.
$$
\begin{array}{l}
\outlying(\rho_0,MSet)\land \textsf{root}=\texttt{r}_0\\
\bigwedge_{x\in \texttt{NS}_0}((\neg x\rightarrow \textsf{mk})\land (\neg x\rightarrow \textsf{chk})\land (x\rightarrow \textsf{l}=\texttt{L}_0(x))
\land (x\rightarrow \textsf{r}=\texttt{R}_0(x)))\\
\mbox{}\ \ \ \ \ \ \{\mbox{the program}\}\\
\rho\land \bigwedge_{x\in \texttt{NS}_0}(x\rightarrow \textsf{mk} \land x\rightarrow \textsf{chk}\land (x\rightarrow \textsf{l}=\texttt{L}_0(x)\land (x\rightarrow \textsf{r}=\texttt{R}_0(x))
\end{array}
$$

\subsection{The proof}
The main part of the program under verification is a while-statement. The invariant of this while-statement is as follow.
$$
\begin{array}{l}
\outlying(\rho_0,MSet)\land\textsf{Acyclic}(\textsf{Stack}(\textsf{p}))\land \textsf{OutPath}(\texttt{NS}_0-\textsf{rng}(\textsf{Stack}(\textsf{p})))\land\\
\textsf{InPath}(\textsf{p})\land (\textsf{NextNode}(\textsf{p})=\textsf{t})
\end{array}
$$
In the rest of this paper, we use $\textrm{SW\_INV}$ to denote this formula. The framework of our proof is depicted in Figure~\ref{SW-ALG-WITH-PROOF}. In the rest part of this section, we will prove the assertions about the sequential statements in line (7), (9) and (11).

\begin{figure}
\begin{tabbing}
\ \ \ \ \ \ \=\ \ \ \ \=\ \ \ \ \ \=\ \ \ \ \ \=\ \ \ \ \ \=\ \ \ \ \ \=\\
    \>$\{\outlying(\rho_0,MSet)\land(\textsf{root}=\texttt{r}_0)\land $\\
    \>$\ \ \bigwedge_{x\in \texttt{NS}_0}((\neg x\rightarrow \textsf{mk})\land (\neg x\rightarrow \textsf{chk})\land (x\rightarrow \textsf{l}=\texttt{L}_0(x))\land (x\rightarrow \textsf{r}=\texttt{R}_0(x)))\}$\\

(1) \>$\textsf{t} := \textsf{root};$\\
(2) \>$\textsf{p} := \textbf{nil};$\\
    \>$\{\outlying(\rho_0,MSet\}\land \textsf{t}=\textsf{root} \land \textsf{p}=\textbf{nil}\}$\\
$\ \ \Rightarrow$ \>$\{\textrm{SW\_INV}\}$\\

(3) \>$\textbf{while }( \textsf{CONDITION}_{wh}  )\textbf{ do } \mbox{//stack not empty or \textsf{t} not explored;}$\\
(4) \>$\textbf{begin}$\\
    \>  \>$\{\textrm{SW\_INV}\land \textsf{CONDITION}_{wh}\}$\\
(5)	\>  \>$\textbf{if }(\textsf{CONDITION}_{1}) \mbox{//t is already explored or t is \textbf{nil};}$\\
(6)	\>  \>$\textbf{then}$\\
    \>  \>  \>$\{\textrm{SW\_INV}\land \textsf{CONDITION}_{wh}\land \textsf{CONDITION}_{1}\}$\\
(7) \>  \>  \>$\textbf{ if }$\=$(\textsf{CONDITION}_2)\textbf{ then}$  //POP;\\
    \>  \>  \>  \>$\{\textrm{SW\_INV}\land \textsf{CONDITION}_{wh}\land \textsf{CONDITION}_{1}\land\textsf{CONDITION}_2\}$\\
(8)	\>  \>  \>  \>\textbf{begin }$\textsf{q}:=\textsf{t};$\ \ \ $\textsf{t}:=\textsf{p};$\ \ \ $\textsf{p} := \textsf{p}\rightarrow\textsf{r};$\ \ $\textsf{t}\rightarrow\textsf{r} := \textsf{q};$\textbf{ end}\\
    \>  \>  \>  \>$\{\textrm{SW\_INV}\}$\\
(9)	\>  \>  \>\textbf{else} //TRY right neighbor;\\
    \>  \>  \>  \>$\{\textrm{SW\_INV}\land \textsf{CONDITION}_{wh}\land \textsf{CONDITION}_{1}\land\neg\textsf{CONDITION}_2\}$\\
(10)	\>  \>  \>  \>\textbf{begin }\\
(11)\>  \>  \>  \>  \>$\textsf{q}:=\textsf{t}$;\ \ \ $\textsf{t}:=\textsf{p}\rightarrow\textsf{r};$\ \ $\textsf{p}\rightarrow\textsf{r} := \textsf{p}\rightarrow\textsf{l};$\\
(12)\>  \>  \>  \>  \>$\textsf{p}\rightarrow\textsf{l} := \textsf{q}; $\ \ \ $\textsf{p}\rightarrow\textsf{chk} := \textbf{true};$\\
(13)\>  \>  \>  \>\textbf{end}\\
    \>  \>  \>  \>$\{\textrm{SW\_INV}\}$\\
    \>  \>  \>$\{\textrm{SW\_INV}\}$\\
(14)\>  \>\textbf{else}	//$\textsf{t}$ is not explored.  PUSH $\textsf{t}$ into the stack;\\
    \>  \>  \>$\{\textrm{SW\_INV}\land \textsf{CONDITION}_{wh}\land \neg\textsf{CONDITION}_{1}\}$\\
(15)\>	\>   \>\textbf{begin }\\
(16)\>  \>  \>  \>$\textsf{q} := \textsf{p};$ \ $\textsf{p} := \textsf{t};$ \ $\textsf{t} := \textsf{t}\rightarrow\textsf{l};$\ \ $\textsf{p}\rightarrow\textsf{l} := \textsf{q};$\\
(17)\>  \>  \>  \>$\textsf{p}\rightarrow\textsf{mk} := \textbf{true};$\ \ $\textsf{p}\rightarrow\textsf{chk} := \textbf{false};$\\
(18)\>  \>  \>\textbf{end}\\
    \>  \>  \>$\{\textrm{SW\_INV}\}$\\
    \>  \>$\{\textrm{SW\_INV}\}$\\
(19)\>\textbf{end}\\
    \>$\{\textrm{SW\_INV}\land \neg \textsf{CONDITION}_{wh}\}$\\
$\Rightarrow$\>$\rho\land \bigwedge_{x\in \texttt{NS}_0}(x\rightarrow \textsf{mk} \land x\rightarrow \textsf{chk}\land (x\rightarrow \textsf{l}=\texttt{L}_0(x)\land (x\rightarrow \textsf{r}=\texttt{R}_0(x))$
\end{tabbing}
\caption{The Schorr-Waite Algorithm with Proof}\label{SW-ALG-WITH-PROOF}
\end{figure}

\subsection{The assertions about the assignment sequences}
\subsubsection{The assignments in line (8).}
Let $\texttt{EXPAND}_1$ be the following formula.
$$
\begin{array}{l}
\outlying(\rho_0,MSet)\land \textsf{Acyclic}(\textsf{Stack}(\overleftarrow{\textsf{p}\rightarrow\textsf{r}})) \land
\textsf{InPath}(\overleftarrow{\textsf{p}\rightarrow \textsf{r}})\land\\
 (\textsf{NextNode}(\overleftarrow{\textsf{p}\rightarrow \textsf{r}})=\overleftarrow{\textsf{p}})\land \textsf{OutPath}(\texttt{NS}_0-\{\overleftarrow{\textsf{p}}\}-\textsf{rng}(\textsf{Stack}(\overleftarrow{\textsf{p}\rightarrow\textsf{r}})))\land\\
\textsf{Explored}(\overleftarrow{\textsf{t}})\land \textsf{Explored}(\texttt{L}_0(\overleftarrow{\textsf{p}}))\land \overleftarrow{\textsf{t}}=\texttt{R}_0(\overleftarrow{\textsf{p}})\land \overleftarrow{\textsf{p}}\rightarrow \textsf{mk}
\end{array}
$$
Form the definition of memory scope terms, and the properties about over-approximation of $\mathfrak{M}(\textsf{InPath})$ and $\mathfrak{M}(\textsf{OutPath})$,

$$\begin{array}{rl}
\overline{\texttt{EXPAND}_1}\Rightarrow &(\overline{\mathfrak{M}(\texttt{EXPAND}_1)}\cap MSet)\subseteq\\
                                        &\{\texttt{Block}(x) | x\in \texttt{NS}_0-\{{\textsf{p}}\}\}\cup \textsf{MarkedFldMk}\cup \{\&{\textsf{p}}\rightarrow \textsf{mk}\}\cup\\ &({\textsf{t}}=\textbf{nil}?\emptyset: \{\&{\textsf{t}}\rightarrow \textsf{chk}\} )\cup
(\texttt{L}_0({\textsf{p}})=\textbf{nil}?\emptyset: \{\&{\texttt{L}_0(\textsf{p}})\rightarrow \textsf{chk}\})\\
\end{array}$$
So we have $\overline{\mathfrak{M}(\texttt{EXPAND}_1)}\cap \{\&\textsf{q}, \&\textsf{t}, \&\textsf{p}, \&\textsf{p}\rightarrow\textsf{r}\}=\textbf{nil}$, thus
$$
\begin{array}{rl}
	&\{\textrm{SW\_INV}\land \textsf{CONDITION}_{wh}\land \textsf{CONDITION}_{1}\land\textsf{CONDITION}_2\}\\
\Rightarrow	&\texttt{EXPAND}_1\land (\texttt{p}\neq \textbf{nil})\\
	&\ \ \ \uparrow \rho_1\rightsquigarrow \overline{\texttt{EXPAND}_1}\\
	&\outlying(\rho_1, \{\&\textsf{q}, \&\textsf{t}, \&\textsf{p}, \&\textsf{p}\rightarrow\textsf{r}\})\land (\textsf{p}\neq \textbf{nil})\\
	&\mbox{}\ \ \ \ \ \ \ \textbf{begin}\ \ \textsf{q}:=\textsf{t};\ \ \ \textsf{t}:=\textsf{p};\ \ \ \textsf{p} := \textsf{p}\rightarrow\textsf{r};\ \ \textsf{t}\rightarrow\textsf{r} := \textsf{q}; \ \ \textbf{ end}\\
	&\rho_1\land(\textsf{q}=\overleftarrow{\textsf{t}})\land (\textsf{t}=\overleftarrow{\textsf{p}})\land (\textsf{p}=\overleftarrow{\textsf{p}\rightarrow\textsf{r}}) \land (\textsf{t}\rightarrow\textsf{r} = \overleftarrow{\textsf{t}})\\
	&\ \ \ \downarrow \rho_1\rightsquigarrow \texttt{EXPAND}_1\\
	&\texttt{EXPAND}_1\land (\textsf{t}=\overleftarrow{\textsf{p}})\land (\textsf{p}=\overleftarrow{\textsf{p}\rightarrow\textsf{r}}) \land (\textsf{t}\rightarrow\textsf{r} = \overleftarrow{\textsf{t}})\\
\Rightarrow &\{\textrm{SW\_INV}\}
\end{array}$$

\subsubsection{The assignments in line (10-13).}
Let $\textsf{EXPAND}_2$ be the formula
$$
\begin{array}{l}
\outlying(\rho_0,MSet)\land\textsf{Acyclic}(\textsf{Stack}(\overleftarrow{\textsf{p}\rightarrow\textsf{r}}))\land\\ \textsf{OutPath}(\texttt{NS}_0-{\textsf{p}}-\textsf{rng}(\textsf{Stack}(\overleftarrow{\textsf{p}\rightarrow\textsf{l}})))\land
\textsf{InPath}(\overleftarrow{\textsf{p}\rightarrow \textsf{l}})\land\\
(\textsf{NextNode}(\overleftarrow{\textsf{p}\rightarrow \textsf{l}})={\textsf{p}})\land
(\overleftarrow{\textsf{t}} = \texttt{L}_0(\textsf{p}))\land (\overleftarrow{\textsf{p}\rightarrow \texttt{r}} = \texttt{R}_0(\textsf{p}))
\end{array}
$$
We have
$$\begin{array}{rl}
\overline{\textsf{EXPAND}_2}\Rightarrow& (\overline{\mathfrak{M}(\textsf{EXPAND}_2)}\cap MSet)\subseteq\\
                            &\{\texttt{Block}(x)|x\in \texttt{NS}_0-\{\textsf{p}\}\}\cup\{\&\textsf{p}\rightarrow \textsf{mk}, \&\textsf{p}\}\\
                            \Rightarrow &\overline{\mathfrak{M}(\textsf{EXPAND}_2)}\cap \{\&\textsf{q}, \&\textsf{t}, \&\textsf{p}\!\rightarrow\!\textsf{l}, \&\textsf{p}\!\rightarrow\! \textsf{r}, \&\textsf{p}\rightarrow\! \textsf{chk}\} = \emptyset  \end{array}$$
Thus we have the following proof,
\begin{tabbing}
\ \ \ \ \ \ \=\ \ \ \ \=\ \ \ \ \ \=\ \ \ \ \ \=\ \ \ \ \ \=\ \ \ \ \ \=\\
    \>$\{\textrm{SW\_INV}\land (\textsf{CONDITION}_{wh}) \land (\textsf{CONDITION}_1)\land \neg(\textsf{CONDITION}_2)\}$\\
$\Rightarrow$\> $\{\textsf{EXPAND}_2\land (\textsf{p}\neq \textbf{nil}) \}$\\
    \>  \>$\uparrow \rho_2\rightsquigarrow \overline{\textsf{EXPAND}_2}$\\
    \>$\{\outlying(\rho_2,\{\&\textsf{q}, \&\textsf{t}, \&\textsf{p}\rightarrow \textsf{l}, \&\textsf{p}\rightarrow \textsf{r}, \&\textsf{p}\rightarrow \textsf{chk}\})\land (\textsf{p}\neq \textbf{nil})\}$\\
(10)\>\textbf{begin }\\
(11)\>  \>$\textsf{q}:=\textsf{t}$;\ \ \ $\textsf{t}:=\textsf{p}\rightarrow\textsf{r};$\ \ $\textsf{p}\rightarrow\textsf{r} := \textsf{p}\rightarrow\textsf{l};$\\
(12)\>  \>$\textsf{p}\rightarrow\textsf{l} := \textsf{q}; $\ \ \ $\textsf{p}\rightarrow\textsf{chk} := \textbf{true};$\\
(13)\>\textbf{end}\\
    \>$\{\rho_2\land (\textsf{q}=\overleftarrow{\textsf{t}})\land (\textsf{t}=\overleftarrow{\textsf{p}\rightarrow\textsf{r}})\land (\textsf{p}\rightarrow\textsf{r}= \overleftarrow{\textsf{p}\rightarrow\textsf{l}})\}\land (\textsf{p}\rightarrow\textsf{l} = \overleftarrow{\textsf{t}})\land$\\
    \>$\ \ (\textsf{p}\rightarrow\textsf{chk} = \textbf{true} )\}$\\
    \>  \>$\uparrow \rho_2\rightsquigarrow \textsf{EXPAND}_2$\\
    \>$\{\textsf{EXPAND}_2 \land (\textsf{q}=\overleftarrow{\textsf{t}})\land (\textsf{t}=\overleftarrow{\textsf{p}\rightarrow\textsf{r}})\land (\textsf{p}\rightarrow\textsf{r}= \overleftarrow{\textsf{p}\rightarrow\textsf{l}})\}\land (\textsf{p}\rightarrow\textsf{l} = \overleftarrow{\textsf{t}})\land$\\
    \>$\ \ (\textsf{p}\rightarrow\textsf{chk} = \textbf{true} )\}$\\
$\Rightarrow$ \>$\{\textrm{SW\_INV}\}$
\end{tabbing}

\subsubsection{The assignments in line (11)}
Let $\textsf{EXPAND}_3$ be the formula
$$
\begin{array}{l}
\outlying(\rho_0,MSet)\land\textsf{OutPath}(\texttt{NS}_0-\overleftarrow{\textsf{t}}-\textsf{rng}(\textsf{Stack}(\overleftarrow{\textsf{p}})))\land
\textsf{InPath}(\overleftarrow{\textsf{p}})\land \\
(\textsf{NextNode}(\overleftarrow{\textsf{p}})=\overleftarrow{\textsf{t}})\land
\overleftarrow{\textsf{t}\rightarrow \textsf{l}}=\texttt{L}_0(\overleftarrow{\textsf{t}}) \land \overleftarrow{\textsf{t}\rightarrow \textsf{r}} = \texttt{R}_0(\overleftarrow{\textsf{t}})
\end{array}$$
Because
$$\begin{array}{rl}
\overline{\textsf{EXPAND}_3}\Rightarrow & (\overline{\mathfrak{M}(\textsf{EXPAND}_3)}\cap MSet) \subseteq \\
                                        & \{\texttt{Block}(x)|x\in \texttt{NS}_0-\textsf{t}\}\cup (\textsf{p}=\textbf{nil}? \emptyset :\{ \textsf{p}\rightarrow \textsf{mk}\})\\
                            \Rightarrow & \overline{\mathfrak{M}(\textsf{EXPAND}_3)}\cap \{\&\textsf{q}, \&\textsf{p}\}\cup \texttt{Block}(\&\textsf{t})= \emptyset
\end{array}$$
Thus
\begin{tabbing}
\ \ \ \ \ \ \=\ \ \ \ \=\ \ \ \ \ \=\ \ \ \ \ \=\ \ \ \ \ \=\ \ \ \ \ \=\\
    \>$\{\textrm{SW\_INV} \land (\textsf{CONDITION}_{wh}) \land \neg (\textsf{CONDITION}_1)\}$\\
$\Rightarrow$   \>$\{(\textsf{t}\neq \textbf{nil})\land \overline{\textsf{EXPAND}_3}\}$\\
    \>  \>$\uparrow \rho_3\rightsquigarrow \overline{\textsf{EXPAND}_3}$\\
    \>$\outlying(\rho_3,\{\&\textsf{q}, \&\textsf{p}\}\cup \texttt{Block}(\&\textsf{t}))\land(\textsf{t}\neq \textbf{nil})$\\
(15)\>\textbf{begin }\\
(16)\>  \>$\textsf{q} := \textsf{p};$ \ $\textsf{p} := \textsf{t};$ \ $\textsf{t} := \textsf{t}\rightarrow\textsf{l};$\ \ $\textsf{p}\rightarrow\textsf{l} := \textsf{q};$\\
(17)\>  \>$\textsf{p}\rightarrow\textsf{mk} := \textbf{true};$\ \ $\textsf{p}\rightarrow\textsf{chk} := \textbf{false};$\\
(18)\>\textbf{end}\\
    \>$\{\rho_3\land (\textsf{q} = \overleftarrow{\textsf{p}})\land (\textsf{p} = \overleftarrow{\textsf{t}})\land (\textsf{t} = \overleftarrow{\textsf{t}\rightarrow\textsf{l}})\land (\textsf{p}\rightarrow\textsf{l} = \overleftarrow{\textsf{p}})\land$\\
    \>$\ \ (\textsf{p}\rightarrow\textsf{mk} = \textbf{true})\land(\textsf{p}\rightarrow\textsf{chk} = \textbf{false})\}$\\
    \>  \>$\downarrow \rho_3\rightsquigarrow {\textsf{EXPAND}_3}$\\
    \>$\{(\textsf{q} = \overleftarrow{\textsf{p}})\land (\textsf{p} = \overleftarrow{\textsf{t}})\land (\textsf{t} = \overleftarrow{\textsf{t}\rightarrow\textsf{l}})\land
(\textsf{p}\rightarrow\textsf{l} = \overleftarrow{\textsf{p}})\land(\textsf{p}\rightarrow\textsf{mk} = \textbf{true})\land$\\
    \>$\ \ (\textsf{p}\rightarrow\textsf{chk} = \textbf{false}) \land \textsf{EXPAND}_3\}$\\
$\Rightarrow$\>$\{\textrm{SW\_INV}\}$
\end{tabbing}

\section{Yet another example: selection-sort program with indirectly specified code}\label{SEC-SEL-SORT}
In this section, we verify an implementation of the selection-sort algorithm. This example shows that Scope Logic can verify
a template, which is a program with a implicitly specified expression.

\subsection{the program}
\subsubsection{Variables}
This program has five variables:
four variables $\textsf{tmp}$, $\textsf{i}$, $\textsf{j}$, and $\textsf{k}$ with type \textbf{int}, and the variable $\textsf{a}$ with type $\textbf{ARR}(\textbf{int}, 100)$.

The boolean expression $order$ appeared in this program is an implicitly specified expression containing two free variables $x_1$ and $x_2$.  This expression satisfies the following conditions
\begin{enumerate}
\item $\forall x,y\cdot order[x/x_1][y/x_2] \lor order[y/x_1][y/x_2]$;
\item $\forall x,y,z\cdot order[x/x_1][y/x_2]\land order[y/x_1][z/x_2]\Rightarrow order[x/x_1][z/x_2]$;
\item $\forall x,y\cdot order[x/x_1][y/x_2] \land order[y/x_1][x/x_2]\Rightarrow x=y$;
\item $\mathcal{M}(order)\cap MSet =\emptyset$, where $MSet$ is  $\{\&\textsf{tmp}, \&\textsf{i}, \&\textsf{j}, \&\textsf{k}\}\cup \texttt{Block}(\&\textsf{a})$.
\item $FV(\mathcal{M}(order))=\emptyset$
\end{enumerate}
The expression specifies a total order. Further more, the memory scope term of $order$ contains no free variables, and is disjoint with
the memory units that may be modified by this program.

\begin{figure}
\begin{tabbing}
\mbox{}\ \ \ \ \ \ \=\ \ \ \ \=\ \ \ \ \=\\
Variables:\\
$\textsf{i}, \textsf{j}, \textsf{k}, \textsf{tmp}$ declared with type $\textbf{int}$.\\
$\textsf{a}$ declared with type $\textbf{ARR}(\textbf{int} ,100)$\\
\\

(1) \>$\textsf{i} := 0;$\\
(2) \>$\textbf{while }( \textsf{i} < 100 )\textbf{ do }$\\
(3) \>$\textbf{begin}$\\
(4) \>  \>$\textsf{j} := \textsf{i}+1;\ \ \textsf{k}:= \textsf{i};$\\
(5) \>  \>$\textbf{while }(\textbf{j} < 100 ) \textbf{ do }$\\
(6) \>  \>$\textbf{begin}$\\
(7) \>  \>  \>$ \textbf{if } order(\textsf{a}[\textsf{j}], \textsf{a}[\textsf{k}])\textbf{ then } \textsf{k} := \textsf{j}; \textbf{\ else skip};$\\
(8) \>  \>  \>$\textsf{j}:=\textsf{j}+1$;\\
(9) \>  \>$\textbf{end}$\\
(10)\>  \>$\textsf{tmp} := \textsf{a}[\textsf{i}];\ \ \textsf{a}[\textsf{i}] := \textsf{a}[\textsf{k}];\ \ \textsf{a}[\textsf{k}]:=\textsf{tmp};\ \ \textsf{i}:=\textsf{i}+1$;\\
(11)\>\textbf{end}\\
\end{tabbing}
\caption{The Selection-Sort Algorithm}\label{SORT-PROGRAM}
\end{figure}

\subsection{The Recursive Functions and Their Memory Scope Functions}
The recursive functions used in this proof are depicted in Figure~\ref{SelSortRecursiveFuns}.
$\textsf{OrderSet}(x,y)$ means that $x$ is less than or equal to all the elements in $y$ w.r.t. $order$.
The function $\textsf{OrderSetSet}(s_1, s_2)$ means that all elements in $s_1$ is less than all elements
in $s_2$ w.r.t. $order$. The function $\textsf{DataSet}(l, r)$ represents the set of elements indexed
from $l$ to $r$ in $\textsf{a}$. $\textsf{AllBut}(l, r, i, j)$ represents the set of elements indexed
from $l$ to $r$ in the array $\textsf{a}$ except that $\textsf{a}[i]$ and $\textsf{a}[j]$ are excluded.
The function $\textsf{arrLocs}(l, r)$ computes the set memory units used by the array elements index
from $l$ to $r$. The corresponding simplified memory scope functions (or over-approximation) are depicted
in Figure~\ref{SelSortMemScope}.

\begin{figure}
\begin{tabbing}
$\textsf{OrderSet}$\=$(x:\textbf{int}, s:\textbf{SetOf}(\textbf{int})):\textbf{bool}\triangleq \bigwedge_{y\in s} order(x,y)$\\
$\textsf{OrderSetSet}(s_1:\textbf{SetOf}(\textbf{int}), s_2:\textbf{SetOf}(\textbf{int})):\textbf{bool}\triangleq \bigwedge_{y\in s_1} \textsf{OrderSet}(y,s_2)$\\
$\textsf{Sorted}(l:\textbf{int}, r:\textbf{int}):\textsf{bool}$\\
        \>$\triangleq (r\le l)?\textbf{true} : order(\textsf{a}[l], \textsf{a}[l+1])\land \textsf{Sorted}(l+1,r)$\\
$\textsf{DataSet}(l:\textbf{int}, r:\textbf{int}):\textbf{SetOf}(\textbf{int})$\\
        \>$\triangleq (r<l)? \emptyset : \{a[l]\}\cup \textsf{DataSet}(l+1,r)$\\
$\textsf{AllBut}(l:\textbf{int}, r:\textbf{int}, i:\textbf{int}, j:\textbf{int}): \textbf{SetOf}(\textbf{int})$\\
        \>$\triangleq (r<l)? \emptyset :((l=i\lor l=j)? \emptyset: \{a[i]\})\cup \textsf{AllData}(l+1,r)$\\
$\textsf{arrLocs}(l:\textbf{int}, r:\textbf{int}):\textbf{SetOf}(\textbf{Ptr})$\\
        \>$\triangleq (r<l)? \emptyset: \{\&a[l]\}\cup \textsf{arrLocs}(l+1,r)$
\end{tabbing}
\caption{The recursive functions used to prove the selection-sort algorithm}\label{SelSortRecursiveFuns}
\end{figure}

\begin{figure}
\begin{center}
\begin{tabular}{|c|c|}
\hline
Recursive functions                 & Memory Scope functions (or over-aproximation)\\
\hline
\hline
$\textsf{OrderSet}, \textsf{OrderSetSet}$       & $\mathfrak{M}(order)$\\
\hline
$\textsf{arrLocs}$   & $\emptyset$\\
\hline
$\textsf{Sorted}(l,r), \textsf{DataSet}(l,r)$                     & $\textsf{arrLocs}(l,r)$\\
\hline
$\textsf{AllBut}(l,r,i,j)$                                & $\textsf{arrLocs}(l,r)-\{\&\textsf{a}[i], \&\textsf{a}[j]\}$\\
\hline
\end{tabular}
\end{center}
\caption{The memory scope functions or over-approximations}\label{SelSortMemScope}
\end{figure}

\begin{figure}
\begin{tabbing}

\end{tabbing}
\caption{The properties used in the proof}
\end{figure}

\subsection{The Specification}
This is an implementation of the selection-sort algorithm.
This algorithm sort the integers in the array. It modifies only the memory units in $MSet$.
So the specification is as follow.
$$\begin{array}{l}
\outlying(\rho_0, MSet)\land S_0 = \textsf{DataSet}(0,99)\\
\mbox{}\ \ \ \ \ \ \{\mbox{The program}\}\\
\outlying(\rho_0,MSet)\land S_0 = \textsf{DataSet}(0,99)\land \textsf{Sorted}(0,99)
\end{array}$$

\subsection{The Proof in Code}
The invariant of the outer while-statement, denoted as $\texttt{INV}_{o}$, is as follow.
$$\begin{array}{l}
\outlying(\rho_o, MSet)\land (\textsf{i}\le 100)\land\\
\textsf{Sorted}(0,{\textsf{i}}-1)\land \textsf{OrderSetSet}(\textsf{DataSet}(0,{\textsf{i}}-1),\textsf{DataSet}({\textsf{i}},99))\land\\
S_0=\textsf{DataSet}(0,99)
\end{array}$$
$\mathfrak{M}(\texttt{INV}_{o})$ is $$\mathfrak{M}(\rho_o)\cup \{\&\textsf{i}\}\cup \mathfrak{M}(order)\cup \textsf{arrLocs}(0,99)$$
The invariant of the inner while-statement, denoted as $\texttt{INV}_{i}$, is as follow.
$$\begin{array}{l}\outlying(\rho_i, \{\&\textsf{k},\&\textsf{j}\})\land\\
\textsf{OrderSet}(\textsf{a}[{\textsf{k}}],\textsf{DataSet}(\textsf{i},{\textsf{j}}-1)) \land (0\le \textsf{i}\le {\textsf{k}}<100)\land (0\le {\textsf{j}}\le 100)
\end{array}$$
%
Let \texttt{EXP} be the formula
$$\begin{array}{l}
\outlying(\rho_0, MSet)\land \textsf{Sorted}(0,\overleftarrow{\textsf{i}}-1)\land\\
\textsf{OrderSetSet}(\textsf{DataSet}(0,\overleftarrow{\textsf{i}}-1), \textsf{AllBut}(\overleftarrow{\textsf{i}},99,\overleftarrow{\textsf{i}},\textsf{k})\cup\{\overleftarrow{\textsf{a}[\textsf{i}]},\overleftarrow{\textsf{a}[\textsf{k}]}   \})\land \\ \textsf{OrderSet}(\overleftarrow{\textsf{a}[\textsf{k}]},\textsf{AllBut}(\overleftarrow{\textsf{i}},99,\overleftarrow{\textsf{i}},\textsf{k}))\land order(\overleftarrow{\textsf{a}[\textsf{k}]},\overleftarrow{\textsf{a}[\textsf{i}]})\land order(\overleftarrow{\textsf{a}[\textsf{k}]},\overleftarrow{\textsf{a}[\textsf{k}]})
\end{array}$$
$\mathfrak{M}(\texttt{EXP})$ is
$$\mathfrak{M}(\rho_o)\cup \{\&\textsf{k}\}\cup \textsf{arrLocs}(0,99)-\{\&\textsf{a}[{\textsf{k}}], \&\textsf{a}[\overleftarrow{\textsf{i}}]  \}  $$

Figure~\ref{SORT-PROGRAM-W-PROOF-1} presents the proof of this program with indirectly specified expression. The following is some properties used in this proof.
$$0\le l<i,j<r\le 99\Rightarrow \textsf{DataSet}(l,r) = \textsf{AllBut}(l,r,i,j)\cup \{\textsf{a}[{\textsf{i}}], \textsf{a}[{\textsf{j}}]\}$$
$$order(x,y)\land \textsf{OrderSet}(x,s)\Rightarrow \textsf{OrderSet}(x,s\cup\{y\})$$
$$\textsf{OrderSetSet}(s_1,s_2)\land \textsf{OrderSet}(x,s_2)\Rightarrow \textsf{OrderSetSet}(s_1\cup\{x\},s_2)$$

\begin{figure}
\begin{tabbing}
\mbox{}\ \ \ \ \ \ \=\ \ \ \ \=\ \ \ \ \=\\
    \>$\{\outlying(\rho_0,MSet)\land S_0=\textsf{DataSet}(0,99)$\}\\
(1) \>$\textsf{i} := 0;$\\
    \>$\{\outlying(\rho_0,MSet)\land S_0=\textsf{DataSet}(0,99)\land (\textsf{i}=0)\}$\\
    $\Rightarrow$\>$\{\texttt{INV}_{o}\}$\\
(2) \>$\textbf{while }( \textsf{i} < 100 )\textbf{ do }$\\
(3) \>$\textbf{begin}$\\
    \>  \>$\{\texttt{INV}_{o}\land (\textsf{i}<100)\}$\\
(4) \>  \>$\textsf{j} := \textsf{i}+1;\ \ \textsf{k}:= \textsf{i};$\\
    \>  \>$\{\texttt{INV}_{o} \land \textsf{OrderSet}(\textsf{a}[{\textsf{k}}],\textsf{DataSet}(\textsf{i},{\textsf{j}}-1)) \land (0\le \textsf{i}\le {\textsf{k}}<100)\land (0\le {\textsf{j}}\le 100)\}$\\
    \>  \>  \>$\uparrow \rho_i\rightsquigarrow \texttt{INV}_0$\\
    \>  \>$\{\texttt{INV}_{i}\}$\\
(5) \>  \>$\textbf{while }(\textsf{j} < 100 ) \textbf{ do }$\\
(6) \>  \>$\textbf{begin}$\\
    \>  \>  \>$\{\texttt{INV}_{i}\land (\textsf{j} < 100 )\}$\\
(7) \>  \>  \>$ \textbf{if } order(\textsf{a}[\textsf{j}], \textsf{a}[\textsf{k}])\textbf{ then } \textsf{k} := \textsf{j}; \textbf{\ else skip};$\\
    \>  \>  \>$\{\texttt{Outlying}(\rho_i, \{\&\textsf{k},\&\textsf{j}\})\land order(\textsf{a}[\textsf{k}],\textsf{a}[\textsf{j}])\land (0\le \textsf{i}\le {\textsf{k}}<100)\land$\\
    \> \> \>$\ \ \textsf{OrderSet}(\textsf{a}[\textsf{k}], \textsf{DataSet}(\textsf{i},\textsf{j}-1)) \land(\textsf{j} < 100 )\}$\\
(8) \>  \>  \>$\textsf{j}:=\textsf{j}+1$;\\
    \>  \>  \>$\{\texttt{INV}_{i}\}$\\
(9) \>  \>$\textbf{end}$\\
    \>  \>$\{\texttt{INV}_{i}\land\neg(\textsf{j} < 100 )\}$\\
    \>$\Rightarrow$\>$\{\texttt{Outlying}(\rho_i, \{\&\textsf{k},\&\textsf{j}\})\land \textsf{OrderSet}(\textsf{a}[{\textsf{k}}],\textsf{DataSet}(\textsf{i},99))\}$\\
    \>  \>\ \ \ \ \ $\downarrow \rho_i\rightsquigarrow \texttt{INV}_0$\\
    \>  \>$\{\texttt{INV}_{o}\land \textsf{OrderSet}(\textsf{a}[{\textsf{k}}],\textsf{DataSet}(\textsf{i},99)) \land (0\le \textsf{i}\le {\textsf{k}}<100)\}$\\
    \>$\Rightarrow$\> $\{ (0\le \textsf{i}\le \textsf{k}<100)\land \overline{\texttt{EXP}}\}$\\
    \>  \>  \>$\uparrow \rho_2\rightsquigarrow \overline{\texttt{EXP}}$\\
    \>  \>$\{\outlying(\rho_2,\{\&\textsf{i}, \&\textsf{tmp},\&\textsf{a}[\textsf{i}], \&\textsf{a}[\textsf{k}]\})\land (0\le \textsf{i}, \textsf{k}<100)\}$\\
(10)\>  \>$\textsf{tmp} := \textsf{a}[\textsf{i}];\ \ \textsf{a}[\textsf{i}] := \textsf{a}[\textsf{k}];\ \ \textsf{a}[\textsf{k}]:=\textsf{tmp};\ \ \textsf{i}:=\textsf{i}+1$;\\
    \>  \>$\{\rho_2\land (\textsf{a}[\textsf{i-1}] = \overleftarrow{\textsf{a}[\textsf{k}]})\land (\textsf{a}[\textsf{k}]= \overleftarrow{\textsf{a}[\textsf{i}]})\land (\textsf{i}=\overleftarrow{\textsf{i}}+1)\}$\\
    \>  \>  \>$\downarrow \rho_2\rightsquigarrow \texttt{EXP}$\\
    \>  \>$\{(\textsf{a}[\textsf{i-1}] = \overleftarrow{\textsf{a}[\textsf{k}]})\land (\textsf{a}[\textsf{k}]= \overleftarrow{\textsf{a}[\textsf{i}]})\land (\textsf{i}=\overleftarrow{\textsf{i}}+1)\land \texttt{EXP}\}$\\
    \>$\Rightarrow$\>$\{\texttt{INV}_{o}\}$\\
(11)\>\textbf{end}\\
    \>$\{\texttt{INV}_{o}\land \neg(\textsf{i}<100)\}$\\
$\Rightarrow$\>$\{\outlying(\rho_0,MSet)\land S_0=\textsf{DataSet}(0,99)\land \textsf{Sorted}(0,99)\}$
\end{tabbing}
\caption{The Selection-Sort Algorithm with Proof}\label{SORT-PROGRAM-W-PROOF-1}
\end{figure}


\begin{thebibliography}{00}
\bibitem{HOARE-1}
C.A.R. Hoare.
\newblock {An axiomatic basis for computer programming}.
\newblock \emph{Communications of the ACM}, 12(10):576-580 and 583, October 1969


\bibitem{RODNEY}
Rodney M. Burstall.
\newblock{Some techniques for proving correctness of programs which alter data structures.}
\newblock{In \emph{Machine Intelligence} 7, pages 23-50. Edinburgh University Press, Edinburgh, Scoland, 1972}

\bibitem{STEPHEN}
Stephen A. Cook and Derek C. Oppen.
\newblock{An assertion language for data structures.}
\newblock{In Conference Record of 2nd ACM Symposium on Priciples of Programming Languages, pages 160-166. New York, 1975}

\bibitem{JOSEPH}
Joseph M. Morris.
\newblock {A general axiom of assignment; assignment and linked
data structures; a proof of the Schorr-Waite algorithm.}
\newblock {In \emph{Theoretical Foundations of Programming Methodology}} pages 25-51. D. Reidel, Dordrecht, Holland 1982.

\bibitem{SEPLOG}
Jonh C. Reynolds
\newblock{An overview of separation logic.}
\newblock{In proceedings of \emph{Verified Software: Theories, Tools, Experiments 2005},  Zurich, Switzerland, October 10-13, 2005}
\newblock {Revised Selected Papers and Discussions}


\bibitem{SEPEXAMPLE}
Hongseok Yang.
\newblock {An example of local reasoning in BI pointer logic:
The Schorr-Waite graph marking algorithm.}
In Fritz Henglein, John
Hughes, Henning Makholm, and Henning Niss, editors, \emph{SPACE 2001:
Informal Proceedings of Workshop on Semantics, Program Analysis and
Computing Environments for Memory Management,} pages 41¨C68. IT
University of Copenhagen, 2001


\bibitem{LPF}
C.B. Jones and C.A.Middelburg
\newblock{A typed logic of partial functions reconstructed classically.}
\newblock{In \emph{Acta Inform}. 31 5 (1994), pp. 399¨C430}


\end{thebibliography}
\end{document}